\documentclass[final]{elsarticle}
\usepackage[left= 3cm, right=3cm]{geometry}
\usepackage{natbib}

\usepackage{pstricks, pst-plot}	
\usepackage{graphicx} 
\usepackage{wrapfig}  
\usepackage[figuresright]{rotating}

\usepackage{amsmath}
\usepackage{amsthm}
\usepackage{amssymb}
\usepackage{dsfont}
\usepackage{hyperref}
\usepackage{mathalpha}
\usepackage{ragged2e}
\usepackage{xcolor}
\usepackage[english]{babel}
\usepackage{blindtext}
\usepackage{caption}
\usepackage{subcaption}
\usepackage[ruled,vlined]{algorithm2e}
\usepackage{algpseudocode}
\usepackage{float}

\usepackage{booktabs} 
\usepackage{multirow}
\usepackage{pifont}
\usepackage{csquotes}
\newcommand{\xmark}{\ding{55}}%
\newtheorem{theorem}{Theorem}
\newtheorem{lemma}[theorem]{Lemma}

\DeclareMathAlphabet{\mathpzc}{OT1}{pzc}{m}{it}
\DeclareMathOperator*{\argmax}{arg\,max}

\usepackage{bm}
\usepackage{mathtools}

\DeclarePairedDelimiter\floor{\lfloor}{\rfloor}
\usepackage{placeins}

\begin{document}
\begin{frontmatter}%

\title{Distributional Reinforcement Learning for Scheduling of Chemical Production Processes } 

 \author[a]{Max Mowbray}
 \author[a]{Dongda Zhang\corref{cor1}}
 \ead{dongda.zhang@manchester.ac.uk}
 \author[b]{Ehecatl Antonio Del Rio Chanona\corref{cor1}}
 \ead{a.del-rio-chanona@imperial.ac.uk} 
 \cortext[cor1]{Corresponding authors}

 \address[a]{Centre for Process Integration, School of Chemical Engineering and Analytical Science, The University of Manchester, Manchester, M13 9PL, United Kingdom}
 \address[b]{Sargent Centre for Process Systems Engineering, Department of Chemical Engineering, Imperial College London, London, SW7 2AZ, United Kingdom}

\begin{abstract}                
Reinforcement Learning (RL) has recently received significant attention from the process systems engineering and control communities. Recent works have investigated the application of RL to identify optimal scheduling decision in the presence of uncertainty. In this work, we present a RL methodology tailored to efficiently address production scheduling problems in the presence of uncertainty. We consider commonly imposed restrictions on these problems such as precedence and disjunctive constraints which are not naturally considered by RL in other contexts. Additionally, this work naturally enables the optimization of risk-sensitive formulations such as the conditional value-at-risk (CVaR), which are essential in realistic scheduling processes. The proposed strategy is investigated thoroughly in a parallel batch production environment, and benchmarked against mixed integer linear programming (MILP) strategies. We show that the policy identified by our approach is able to account for plant uncertainties in online decision-making, with expected performance comparable to existing MILP methods. Additionally, the framework gains the benefits of optimizing for risk-sensitive measures, and identifies online decisions orders of magnitude faster than the most efficient optimization approaches. This promises to mitigate practical issues and ease in handling realizations of process uncertainty in the paradigm of online production scheduling.
\end{abstract}
\end{frontmatter}

\section{Introduction}
\subsection{Online production scheduling: optimization and simulation}

The development of methods for efficient production scheduling of batch processes is an area of significant interest within the domain of process systems engineering and operations research \cite{sarkis2021decision, kis2020resources}. There are three main drivers for research within online production scheduling: a) identifying modelling approaches that integrate with the practicalities of online scheduling, b) considering plant uncertainties, and c) handling nonlinearities \cite{harjunkoski2014scope}. There is a diverse array of modelling and solution methods for production scheduling problems \cite{maravelias2012general}. On one hand, modelling approaches can be broadly classified as discrete or continuous time \cite{floudas2004continuous}; on the other hand, optimization approaches can be characterized as simulation-based or optimization. The former is generally underpinned by stochastic search algorithms \cite{spall2005introduction}, whereas the latter is dominated via the use of mixed integer programming (MIP). However, industrial practice is generally dominated by the use of heuristics given the challenges outlined in a-c.

Recent works have argued that formulation of the underlying scheduling problem in terms of a discrete-time, state space model alleviates many of the problems associated with the former challenge (i.e. a)) \cite{gupta2017general}. The benefits of this approach primarily relate to the ease of incorporating state-feedback into the scheduling MIP model and mitigate the requirement for various heuristics to update the model when uncertainty is observed (as is the case in other approaches e.g. continuous-time models) \cite{subramanian2012state}. However, the intuitive benefits inherited from state-space modelling produces scheduling models of a much larger size than continuous-time formulations. This has implications for the solution time and suboptimality can be introduced if the discretization of the time domain is too coarse. Further, considering uncertainty within the model formulation is essentially intractable via MIP solution methods, at least for problems of industrial size. For example, in the case of stochastic MIP, considerable approximations to the scenario tree are required to solve the models (even offline) \cite{li2021review}. This means the most practical way to consider uncertainty is often via robust formulations. This leads to smaller model sizes but requires approximation of plant uncertainties via a few finite dimensional, deterministic expressions \cite{ben2009robust}. It is well known that such formulations are generally conservative \cite{li2008robust}. 

Considering uncertainty is widely acknowledged as an important facet of solution approaches \cite{li2008process, beykal2022data, tian2018synthesis}. In the contenxt of scheduling, the simulation community has demonstrated strong benefits in handling uncertainty \cite{oyebolu2019dynamic, fu2019two}. The high level idea here is generally to compute approximately optimal schedules by evaluating their performance via a Monte Carlo method and a stochastic search algorithm. These approaches have also proven useful in integrating decision-making functions where nonlinearities often arise in the underlying model \cite{dias2018simulation}. This is primarily because one avoids the formulation of mixed integer nonlinear programs (MINLP). However, it should be noted that optimization approaches fair well with regard to integrated problems when the resultant formulation is mixed integer linear programming (MILP) \cite{santos2021integrated,charitopoulos2019closed}. In addition, simulation based approaches are generally expensive to conduct online because of the sample inefficiency associated with Monte Carlo methods and stochastic search algorithms. 

The expense of conducting simulations online has led to the development of Reinforcement Learning (RL) based approaches to online production scheduling \cite{hubbs2020deep}. The primary idea here is to exploit the Markov decision process (MDP) framework to model production scheduling systems as a stochastic process. One can then use RL to learn a functionlization of an optimal decision policy for the underlying environment through offline simulations of the model. The policy function can then be deployed online to provide scheduling decisions, instead of online optimization. This allows the scheduling element to a) consider nonlinearities and uncertainties within the process dynamics, b) use arbitrary distributions of the uncertainty, and c) identify scheduling decisions in a very short time frame (i.e. on the order of micro to milliseconds). Despite the additional work required in offline, simulation-based policy learning, these facets are appealing in the context of modern production environments. 

\subsection{Online production scheduling and Reinforcement Learning}

Although RL has been demonstrated for sequential decision making in a number of case studies \cite{caputo2022analyzing,lawrence2022deep,degrave_magnetic_2022, yoo2021reinforcement}, its application to physical production systems has been relatively limited. For example, \cite{waschneck2018optimization} applied deep Q networks to optimize a flexible jobshop (i.e. a multipurpose multi-stage production facility), however, the authors provide little information as to how the method proposed accounts for constraints. Further, the approach is benchmarked to common heuristics rather than optimization formulations. In \cite{palombarini2018generating}, an RL rescheduling approach is presented to repair and ensure the feasibility of the schedule when subject to realizations of plant uncertainty. However, the method does not consider the optimality of the original schedule and is not benchmarked to an existing method. More recently, \cite{hubbs2020deep} provided an extensive analysis of a Reinforcement Learning approach compared to deterministic and stochastic MIP for a single-stage, continuous production scheduling problem. The RL method is demonstrated to be an appealing solution approach. Despite their achievements, as this is a first proof-of-concept, the case study considered is relatively simple and does not involve precedence constraints or requirements for setup time between the end and start of subsequent operations within a given unit. 

In addition, considering constraints is an important step in the development of RL algorithms, given the MDP framework does not provide an explicit mechanism to handle them. There has been much research in other decision-making problems regarding this \cite{petsagkourakis2022chance,achiam2017constrained, zhang2020first, MowbrayM2022Sccr}. The presence of precedence and disjunctive constraints in production scheduling provides a challenge of a different nature. Detailed examples of these constraints are explored in Section \ref{sec:constraint_handling}. Further, the use of RL poses challenges such as robustness and reliability \cite{waubert2022reliability}. 

Recently, there has been a lot of interest in the development of distributional RL algorithms \cite{bdr2022}. Instead of formalizing the objective via expected performance, distributional RL algorithms try to find the optimal policy for other measures of the performance distribution \cite{bdr2022}. This enables identification of more risk-sensitive policies and consideration of the tails of the policy performance \cite{tang2019worst}. The proposition of risk-sensitive formulations has been common to mathematical programming for some time \cite{rockafellar2002conditional}, however, its use in scheduling problems has been limited \cite{najjarbashi2019variability,chang2017distributionally}. In this work, we utilize distributional RL to consider risk-sensitive formulations and low probability, worst-case events. This is particularly important in engineering and business applications, where decision-making is generally risk-averse and catastrophic events are highly disfavoured. 

\subsection{Contribution}

In this work, we develop the methodology of Reinforcement Learning to production scheduling by proposing a novel and efficient policy optimization method that a) handles precedence and disjunctive constraints, b) is practical and observes stability in learning, and c) provides means to identify risk-sensitive policies. In doing so, we improve the reliability and robustness of RL algorithms, but also inherit the advantages of identifying online scheduling decisions in real time from a function, and ease in handling nonlinearity and uncertainty.

To handle a), we present a logic based framework that implements transformations of the RL scheduling decisions to ensure they satisfy those constraints derived from propositional logic (i.e. precedence and disjunctive constraints). This is discussed in detail in Section \ref{sec:constraint_handling}. The optimization of the policy in view of this framework is handled by a stochastic search optimization algorithm (PSO-SA) which combines particle swarm optimization (PSO) and simulated annealing (SA) to balance exploitation and exploration of the policy parameters. The use of stochastic search optimization approaches lend themselves naturally to distributional RL and facilitate b) and c). Specifically, one can estimate risk-sensitive measures such as the conditional value-at-risk via a Monte Carlo method. This has particular advantage over other policy gradient and action-value based approaches to distributional RL, as one is not required to make assumption on the form of the performance distribution induced by the policy. Further, the use of stochastic search methods removes the dependence on first-order policy learning methods, which can lack robustness. This is partly due to the use of noisy directions for policy improvement, whose evaluation is known to be expensive (i.e. policy gradients and deep Q learning) and sometimes unreliable \cite{riedmiller2007evaluation, nota2020policy}.

The proposed method is benchmarked on a classical production scheduling problem against an MIP approach. The problem is a multiproduct batch plant with parallel production lines. The case study is a modified version of the original study provided in \cite{cerda1997mixed} to include processing time and due date uncertainty. Extensive analysis of the results is provided.

The rest of this paper is organized as follows: in Section \ref{sec:PS}, we present the problem statement; in Section \ref{sec:methodology} we outline a novel and efficient RL approach to scheduling; in Section \ref{sec:CS}, we outline the details of a case study, the results and discussion of which are presented subsequently in Section \ref{sec:R&D}; and, in Section \ref{sec:conclusion} we finish with concluding thoughts and plans for future work.

\section{Problem Statement}\label{sec:PS}
The scheduling problem is generally subject to both endogenous and exogenous sources of uncertainty. In this work, we focus on the online scheduling of parallel, sequential batch operations in a chemical production plant \cite{maravelias2012general}. We assume that the state of the plant at a given time index, $t \in \{0, \ldots, T\}$, within a discrete finite time horizon (of length $T$), is represented by a state, $\mathbf{x}_t \in \mathbb{X}\subseteq \mathbb{R}^{n_x}$, where $\mathbf{x}_t$ can be thought as (but not limited to) current product and raw material inventory, unit availability, and tasks currently being processed. At discrete time steps within the scheduling process of the plant, the scheduler (agent or algorithm who decides the scheduling actions) is able to observe the state of the plant, and select a control action, $\mathbf{u}_t \in \mathbb{U}\subseteq\mathbb{Z}^{n_u}$, which represents an appropriate scheduling decision on the available equipment. The state of the plant then evolves according to the following difference equation:
\begin{equation}\label{eq:plantdyn}
    \begin{aligned}
    \mathbf{x}_{t+1} = f(\mathbf{x}_t, \mathbf{u}_t, \mathbf{s}_t)
    \end{aligned}
\end{equation}
where $\mathbf{s}_{t}\in \mathbb{S}\subseteq\mathbb{R}^{n_s}$ represents a realization of some uncertain plant parameters or disturbance. Eq. \ref{eq:plantdyn} describes the stochastic evolution of the plant and could be equivalently expressed as a conditional probability density function (CPDF), $p(\mathbf{x}_{t+1}|\mathbf{x}_{t}, \mathbf{u}_{t})$. Here, we identify that the system has the Markov property (i.e. the future state only depends on the current state and control actions) and hence the system may be described as a Markov decision process (MDP).

The aim of the scheduler is to minimize objectives such as makespan (which defines the time to complete all of the required tasks on the available equipment) and the tardiness of product completion. Given these objectives, one can define a reward function, $R:\mathbb{X} \times \mathbb{U} \times \mathbb{X} \rightarrow \mathbb{R}$, which describes the performance of the decisions taken (e.g. the higher the reward, the  more profit achieved by the scheduler). Solution methods for MDPs aim to identify a control policy, $\pi:\mathbb{X} \rightarrow \mathbb{U}$, whose aim is to maximize the reward:
\begin{subequations}
\begin{equation}
    \begin{aligned}\label{eq:return}
    Z = \sum_{t=0}^{T-1} R_{t+1}
    \end{aligned}
\end{equation}
\begin{equation}\label{eq:expmax}
    \begin{aligned}
    \pi^* = \argmax_\pi \mathbb{E}_{\pi}\big[Z|X_0 \sim p(\mathbf{x}_0)\big]
    \end{aligned}
\end{equation}
\end{subequations}
where $X_0\in \mathbb{X}$ is the initial state of the plant, which is treated as a random variable and described by an initial state distribution,  $p(\mathbf{x}_0)$. The \textit{return}, $Z \sim p_\pi(z)$, is a random variable that is described according to a probability density function, $p_\pi(z)$, under the current policy, $\pi$, because the plant dynamics are subject uncertainty. Generally, exact expressions of $p_\pi(z)$ in closed form are unavailable. Therefore, the solution policy, $\pi^*$, (i.e. Eq. \ref{eq:expmax}) is evaluated via the sample average approximation. 

Operationally, there is a constraint set, $\mathbb{\hat{U}}_t = \hat{\mathbb{U}}_t^{(1)} \times \hat{\mathbb{U}}_t^{(2)} \ldots \times \hat{\mathbb{U}}_t^{(n_u)}  \text{, where } \mathbb{\hat{U}}^{(l)}_t\subset\mathbb{Z}$, that defines the available tasks or jobs that may be scheduled in the $n_u$ units at any given time index. This may be defined by the viable sequencing of operations in units; requirements for unit cleaning and maintenance periods; requirements for orders to be processed in campaigns; and that processing of batches must be finished before another task is assigned to a given unit, to name a few. General intuition behind these constraints is provided in Section \ref{sec:constraint_handling}. These constraints may or may not be functions of uncertain process variables (e.g. if processing times are subject to uncertainties). In essence, we are trying to solve a discrete time finite horizon stochastic optimal control problem (SOCP) of the form:
\begin{equation}
\mathcal{P}(\pi):=\left
\{\begin{aligned}
        &\max_{\pi} \mathbb{E}_{\pi}\big[Z\big]\\
    &\text{s.t.}\\
    & X_0 \sim p(\mathbf{x}_0)\\
    & \mathbf{s}_t \in \mathbb{S}\subseteq \mathbb{R}^{n_s}\\
    &\mathbf{x}_{t+1} = f(\textbf{x}_t, \mathbf{u}_t, \mathbf{s}_t)\\
    &\mathbf{u}_t = \pi(\mathbf{x}_t)\\
    &\mathbf{u}_t\in\mathbb{\hat{U}}_t \subset \mathbb{Z}^{n_u}\\
    &  \forall t \in \left\{0,...,T\right\}\label{eq:CSOCP}
\end{aligned}\right.
\end{equation}
where $\pi:\mathbb{X} \rightarrow \mathbb{U}$ is a control policy, which takes the state as input and outputs a control action, $\mathbf{u}_t$. In practice, mixed integer approaches to the scheduling problem either make large approximations to the SOCP formed (i.e. via stochastic programming) or assume description of a nominal model, neglecting the presence of uncertain variables entirely. The latter approach is especially common when considering problem sizes of industrial relevance.
In the following section, we present an efficient and novel methodology to identify a policy $\pi$, which provides an approximately optimal solution to the finite horizon SOCP detailed by Eq. \ref{eq:CSOCP}, via RL.  Specifically, $\pi(\theta, \cdot)$ is defined by a neural network with parameters, $\theta \in \mathbb{R}^{n_\theta}$. The problem (i.e. Eq. \ref{eq:CSOCP}) is then reduced to identifying the optimal policy parameters, $\theta^*$.
\section{Methodology}\label{sec:methodology}
The approach that follows is proposed given the following characteristics of the RL production scheduling problem: a) Eq. \ref{eq:CSOCP} formulates control inputs (decisions) as discrete integer values, $\mathbb{{U}}\subset \mathbb{Z}^{n_u}$, that identify the allocation of a task (or job) in a unit at a given time index, and b) one must handle the hard constraints imposed by $\mathbf{u}_t \in \mathbb{\hat{U}}_t$.

\subsection{Identifying discrete control decisions}\label{sec:discretecontrols}

In this work, we are concerned with identifying a policy, which is suitable for optimization via stochastic search methods and is able to handle relatively large\footnote{The term \enquote{large} is used to indicate a control space with high cardinality, $\lvert\mathbb{U}\rvert$.} control spaces. The reasoning behind the use of stochastic search is discussed extensively in Section \ref{sec:SSPO}. However, as stochastic search methods are known not to perform well when the effective dimension of a problem is high\footnote{Stochastic search is known to be less effective when the number of decision variables is in the order of $1000$s, as is common in the neural network function approximations often used in RL.}, there is a requirement for careful construction of the policy function. 

We propose to make predictions via the policy function in a continuous latent space, $\mathbf{w}_t\in \mathbb{W}\subseteq \mathbb{R}^{n_u}$, and then transform that prediction to a corresponding discrete control decision, $\mathbf{u}_t$. As a result, the policy function requires an output dimensionality equivalent to that of the control space (i.e. $n_u$ the number of units or equipment items in the context of production scheduling). The transformation could be defined by a stochastic or deterministic rounding policy. For example, the nearest integer function (i.e. round up or down to the nearest integer), denoted $f_r:\mathbb{W}\rightarrow\mathbb{U}$, is a deterministic rounding policy demonstrated implicitly by many of the studies provided in \cite{hubbs2020orgym}. Both of these transformations are non-smooth and generally assume that the latent space, $\mathbf{w}\in \mathbb{W}$, is a relaxed, continuous equivalent of the original control space. In this work, we implement a nearest integer function approach, $f_r(\mathbf{w})=nint(\mathbf{w})$. 

In the context of this work, this approach is appealing because it minimizes the number of output nodes required in the policy function, and as a result, the number of function parameters. This lowers the effective dimensionality of the problem compared to other means of control selection (e.g. directly predicting the probability mass of a control in the output of the function, as is common in policy gradients with discrete control spaces\footnote{This approach requires $\lvert\mathbb{U}\rvert$ nodes in the output layer of the policy function, where $\lvert\mathbb{U}\rvert > > n_u$. }). This is key for the use of stochastic search optimization methods. 


In the subsequent section, we explore the development of an approach to handling constraints imposed on the scheduling problem.
\FloatBarrier
\subsection{Constraint handling}\label{sec:constraint_handling}

Handling the constraints imposed on RL control selection (input constraints) is generally done explicitly \cite{MowbrayM2022Sccr, pan2021constrained}. In the class of problem of concern to this work, the structure of the constraints on the control space arises from standard operating procedures (SOPs). SOPs can generally be condensed into propositional logic. For example, consider the special case of a production environment with one processing unit available and two tasks \textit{i} and \textit{m} that require processing. We may define a sequencing constraint through the following statement: \enquote{task \textit{i} can only be processed after task \textit{m}}. In the language of propositional logic, this statement is associated with a \textit{True} or \textit{False} boolean. If task \textit{m} has just been completed and the statement is \textit{True} (i.e. task \textit{i} can succeed task \textit{m}), then task \textit{i} belongs to the constraint set at the current time index \textit{t}, i.e. $i\in \hat{\mathbb{U}}_t$. This is known as a precedence constraint. More complex expressions can be derived by relating two or more propositional statements. For example, consider the following expression that relates requirements for starting, $T^{s}$, and end times, $T^f$, of tasks $i$ and $m$ in a production environment with one unit: 
\begin{align*}
    \{T^s_i \geq T^f_{m}\} \lor \{T^s_{m} \geq T^f_{i}\}
\end{align*}
where $\lor$ is a disjunction operator and can be interpreted as an OR relation. Essentially, this statement says that no two tasks can be processed in a given unit at the same time, and is otherwise known as a disjunctive constraint \cite{maravelias2012general}. Such scheduling rules are conventionally transcribed into either generalized disjunctive programming or mixed integer programming formulations. Both solution methods enable constraint satisfaction.

In this work, we hypothesise that a set of controls, $\mathbf{u}_t\in\mathbb{\bar{U}}_t\subseteq \mathbb{Z}^{n_u}$, may be identified that adhere to the logic provided by the SOPs at each discrete control interaction, based on the current state of the plant, $\mathbf{x}_t \in \mathbb{X}_t$. This functional transformation is denoted, $f_{SOP}:\mathbb{U}\times\mathbb{X}\rightarrow \mathbb{\bar{U}}$, where $\mathbb{\bar{U}}\subseteq \mathbb{Z}^{n_u}$, and is assumed non-smooth.

If we are able to define and satisfy all constraints via propositional logic (i.e. $f_{SOP}$), we proceed to use the rounding policy, $f_r$, defined in Section \ref{sec:discretecontrols}. Specifically, by redefining $f_r: \mathbb{W}\rightarrow \mathbb{\bar{U}}$, the implementation provides means to identify a mapping from a continuous latent space to (possibly discrete) controls that are feasible. This mechanism is widely known via action shaping, action masking or domain restriction, and has been shown to significantly improve the efficiency of RL learning process \cite{kanervisto2020action, hubbs2020orgym}.

In the case one is unable to identify (and satisfy) all constraints $\mathbb{\hat{U}}$ via $f_{SOP}$, we penalise the violation of those constraints that cannot be handled, and incorporate a penalty function for the constraint violation, $\phi:\mathbb{X} \times \mathbb{U} \times \mathbb{X} \rightarrow \varphi_{t+1}\in \mathbb{R}$. For example, using discrete-time state space models, one cannot impose the constraint \enquote{no two units can start processing the same task, at the same time} via this logical transformation, without imposing considerable bias on the control selection of the policy. This is discussed further in \ref{app:constraintset}. Given $\mathbf{g}(\mathbf{x},\mathbf{u}) = [g_1(\mathbf{x},\mathbf{u}), \ldots,g_{n_g}(\mathbf{x},\mathbf{u})]$, where $g_i:\mathbb{X}\times \mathbb{U}\rightarrow\mathbb{R}$, $\forall i \in \{1, \ldots, n_g\}$, represent the $n_g$ constraints that cannot be handled innately via propositional logic, we define the \textit{penalised return} from an episode as:
\begin{equation}
    \begin{aligned}\label{eq:zphi}
    \phi    &= R - \kappa_g \left\lVert[\mathbf{g}(\mathbf{x},\mathbf{u})]^+\right\rVert_p\\
    Z^{\phi} &= \sum_{t=0}^{T-1} \varphi_{t+1}
    \end{aligned}
\end{equation}
where $\kappa_g \in \mathbb{R}$ is the penalty weight; $[\mathbf{y}]^+ = \max(0, \mathbf{y})$ defines an element wise operation over $\mathbf{y} \in \mathbb{R}^{n_y}$; and, $\left\lVert \cdot \right\rVert_p$ defines the general $l_p$-norm. From here we can identify a solution policy to Eq. \ref{eq:CSOCP}, $\pi^*$ as follows:
\begin{equation}
    \begin{aligned}\label{eq:expectationopt}
    \pi^* &= \argmax_{\pi} \mathbb{E}_{\pi}\big[Z^{\phi}\big] 
    \end{aligned}
\end{equation}
A figurative description of the approach proposed is provided by Fig. 1, and formalized by Algorithm \ref{alg:rollout}. 
\begin{figure}[h]
    \centering
    \includegraphics[scale = 0.35]{{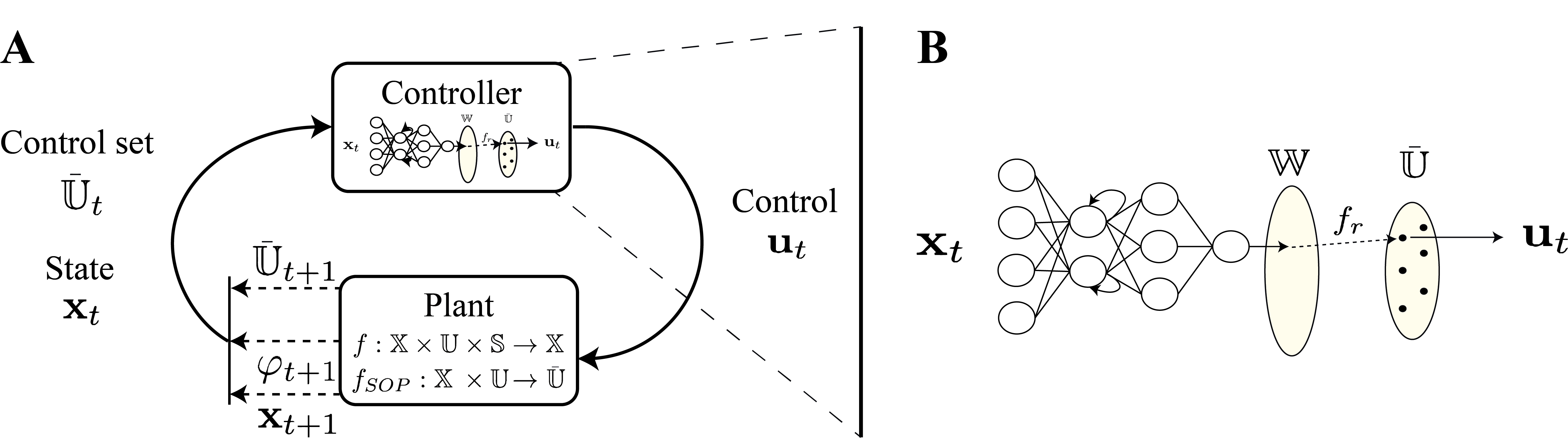}}
    \caption{Figurative description of the feedback control framework used to formulate the scheduling problem. A) A feedback control framework that utilizes logic to identify feasible scheduling decisions. B) Control selection via a deterministic rounding policy.}
    \label{fig:constraint_handling}
\end{figure}

\vspace{0.2cm}
\begin{algorithm}[h]
\SetAlgoLined
\caption{Control selection for production scheduling of an uncertain plant}
\vspace{0.1cm}
\justify
\textbf{Input}: Policy function, $\pi(\theta_0,\cdot)$; functional transformation derived from plant standard operating procedure, $f_{SOP}:\mathbb{X} \times \mathbb{U} \rightarrow \mathbb{\bar{U}}$; description of uncertain plant dynamics, $f: \mathbb{X} \times \mathbb{U} \times \mathbb{S} \rightarrow \mathbb{X}$ (see Eq. \ref{eq:plantdyn}); rounding policy, $f_r:\mathbb{W}\rightarrow\mathbb{\bar{U}}$; initial state distribution, $p(\mathbf{x}_0)$; the number of units (equipment) available, $n_u$; the number of tasks to be processed, $n_T$; the control set, $\mathbb{U} =\{(u^{(1)}, \ldots, u^{(n_u)}) \mid u^{(l)}\in \mathbb{U}^{(l)} \text{, } \forall l \in \{1, \ldots n_u\} \}$, where $\mathbb{U}^{(l)}\subset \mathbb{Z}_+$; penalty function, $\phi:\mathbb{X}\times\mathbb{U}\times\mathbb{X}\rightarrow \mathbb{R}$; and, empty memory buffer, $\mathcal{B}_{info}$\;\vspace{0.1cm}\justify
\textbf{1.} Draw initial state, $\mathbf{x}_0 \sim p(\mathbf{x}_0)$\vspace{0.1cm}\\
\textbf{2.} \For{$t = 0, \ldots, T-1$}{
                    \textbf{a.} Identify control set, $\mathbb{\bar{U}}_t = \{f_{SOP}(\mathbf{x}_t, \mathbf{u}) \in \mathbb{Z}^{n_u}, \forall \mathbf{u} \in \mathbb{U}\} \subset \mathbb{Z}^{n_u}$\vspace{0.1cm}\;
                    \textbf{b.} Predict latent coordinate conditional to current state, $\textbf{w}_t = \pi(\textbf{x}_t;\theta)$\vspace{0.1cm}\;
                    
                    \textbf{c.} Implement rounding policy, $\textbf{u}_{t} = f_r(\mathbf{w}_t)$\vspace{0.1cm}\;
                    
                    \textbf{d.} Implement scheduling decision and simulate, $\mathbf{x}_{t+1} = f(\mathbf{x}_t, \mathbf{u}_t, \mathbf{s}_t)$, where $\mathbf{s}_t \in \mathbb{S}$\vspace{0.1cm}\;
                    
                    \textbf{e.} Observe feedback from penalty function, $\varphi_{t+1} = \phi(\mathbf{x}_t, \mathbf{u}_{t}, \mathbf{x}_{t+1})$\vspace{0.1cm}\;}

\justify\textbf{3.} Assess $Z^{\phi}$ (see Eq.\ref{eq:zphi}) and store, together with information required for policy optimization, in $\mathcal{B}_{info}$
\justify\textbf{Output:}  $\mathcal{B}_{info}$
\label{alg:rollout}
\end{algorithm} 
\subsection{Stochastic search policy optimization}\label{sec:SSPO}
Due to the nature of the problem at hand, this work proposes the use of stochastic search policy optimization algorithms. In general, any stochastic search algorithm can be used. A high level description of the algorithm used in this work is provided by Algorithm \ref{alg:generalSSPO}. Summarizing this approach, we use a hybrid algorithm, which combines particle swarm optimization (PSO) \cite{kennedy1995particle} with simulated annealing (SA) \cite{kirkpatrick1983optimization} and a search space reduction strategy \cite{park2005particle}. For clarity, we refer to this algorithm simply as PSO-SA. The hybridization of the two algorithms helps to efficiently balance exploration and exploitation of the policy function's parameter space. For more details on this algorithm, please refer to \ref{app:PSO-SA}. 

The use of stochastic search policy optimization inherits three major benefits. The first being that stochastic search algorithms are easily parallelizable, which means offline computational time in policy learning can be reduced. The second benefit is that one is freed from reliance on accurate estimation of first order gradients indicative of directions for policy improvement, as in policy gradients and deep Q learning. Estimation of these directions is known to be computationally expensive \cite{riedmiller2007evaluation}, and there is potential for policies to become stuck in local optima, as well as instability in policy learning. This is particularly likely because the loss landscape with respect to the policy parameters is thought to be non-smooth and rough in scheduling problems\footnote{This is due to the structure of the control space and the nature of the scheduling task, i.e. controls that are close in the control space (have small euclidean distance between them) do not necessarily induce similar process dynamics.} \cite{pmlr-v139-merchant21a,amos2022tutorial}. The final bonus is that one can easily optimize for a variety of measures of the \textit{distribution of penalised return}, $p_\pi(z^\phi)$, (i.e. go beyond optimization in expectation as is declared in Eq. \ref{eq:expmax}) provided one is able to obtain a sufficient number of samples. This is discussed further in the next section.

\subsection{Optimizing for the distribution of returns}\label{sec:DistRL}
Recent developments within RL have enabled the field to move beyond optimization in expectation. This subfield is known as \textit{distributional} RL. One of the advantages provided by distributional RL is that one can incentivize optimization of the tails of the distribution of returns, providing a more risk-sensitive formulation. The concept of distributional RL was first introduced by \cite{bellemare2017distributional} in 2017. However, optimization of risk-sensitive criteria has been well established in the portfolio optimization and management community for some time, with the advent of e.g. value-at-risk (VaR), conditional value-at-risk (CVaR) \cite{rockafellar2000optimization}, and Sharpe's ratio. Conventionally, (stochastic gradient-based) distributional RL algorithms are dependent upon making approximations of the probability density function of returns $p_\pi(z^\phi)$, so as to gain tractable algorithms \cite{bellemare2017distributional,tang2019worst}. The major benefit of stochastic search approaches is that one is freed from such approximations by simply estimating the interested measure (e.g. mean, varaince, (C)VaR) directly from samples (i.e. a Monte Carlo method).
In the following, we outline the CVaR in the context of stochastic search policy optimization, and reasons for its use as metric to optimize $p_\pi(z^\phi)$.

\subsubsection{The conditional value-at-risk (CVaR)}
The CVaR is closely related to the value-at-risk (VaR). The VaR defines the value, $z_\beta^\phi$, of the random variable, $Z^\phi$, that occurs with probability less than or equal to a certain pre-defined level, $\beta=[0,1]$, under the cumulative distribution function (CDF), $F_\pi(z^\phi) =\mathbb{P}(Z^\phi \leq z^\phi | \pi)$. Informally, VaR (for some $\beta$) gives us a value, which is the best possible value we can get with a probability of at most $\beta$. For example, conceptualize that $Z^\phi$ represents the mass produced in a process plant per day (kg/day). If the plant has $z_\beta^\phi = 1000$ kg/day, and $\beta=0.05$, this means that there is a 0.05 probability that the production will be of 1000 kg/day or less. 

\begin{figure}[h!]
    \centering
    \subfloat[\centering A probability density function view \vspace{0.1cm}]{{\includegraphics[scale=0.17]{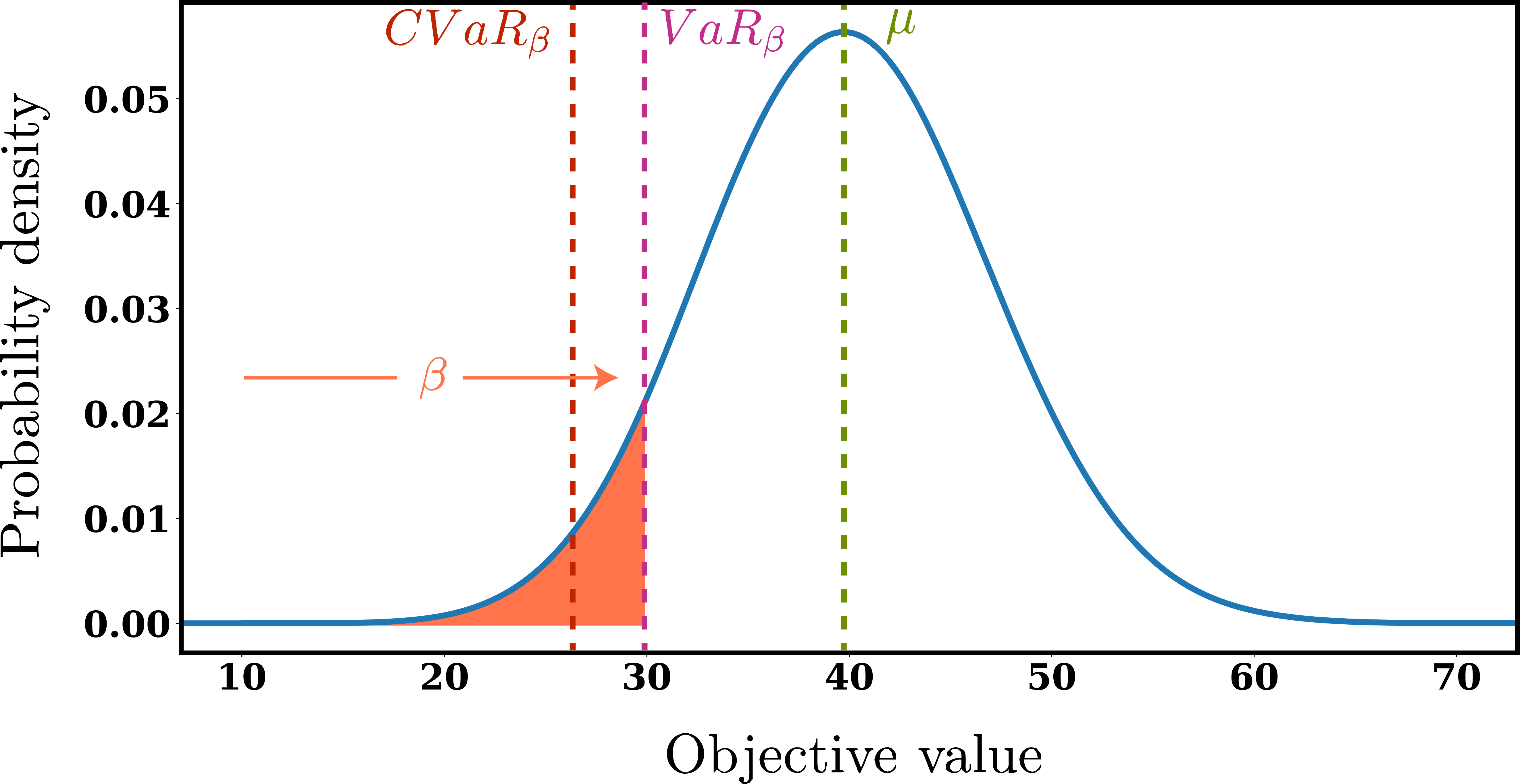}}\label{fig:cvarpdf}}%
    \quad
    \centering
    \subfloat[\centering A cumulative distribution function view \vspace{0.1cm}]{{\includegraphics[scale=0.17]{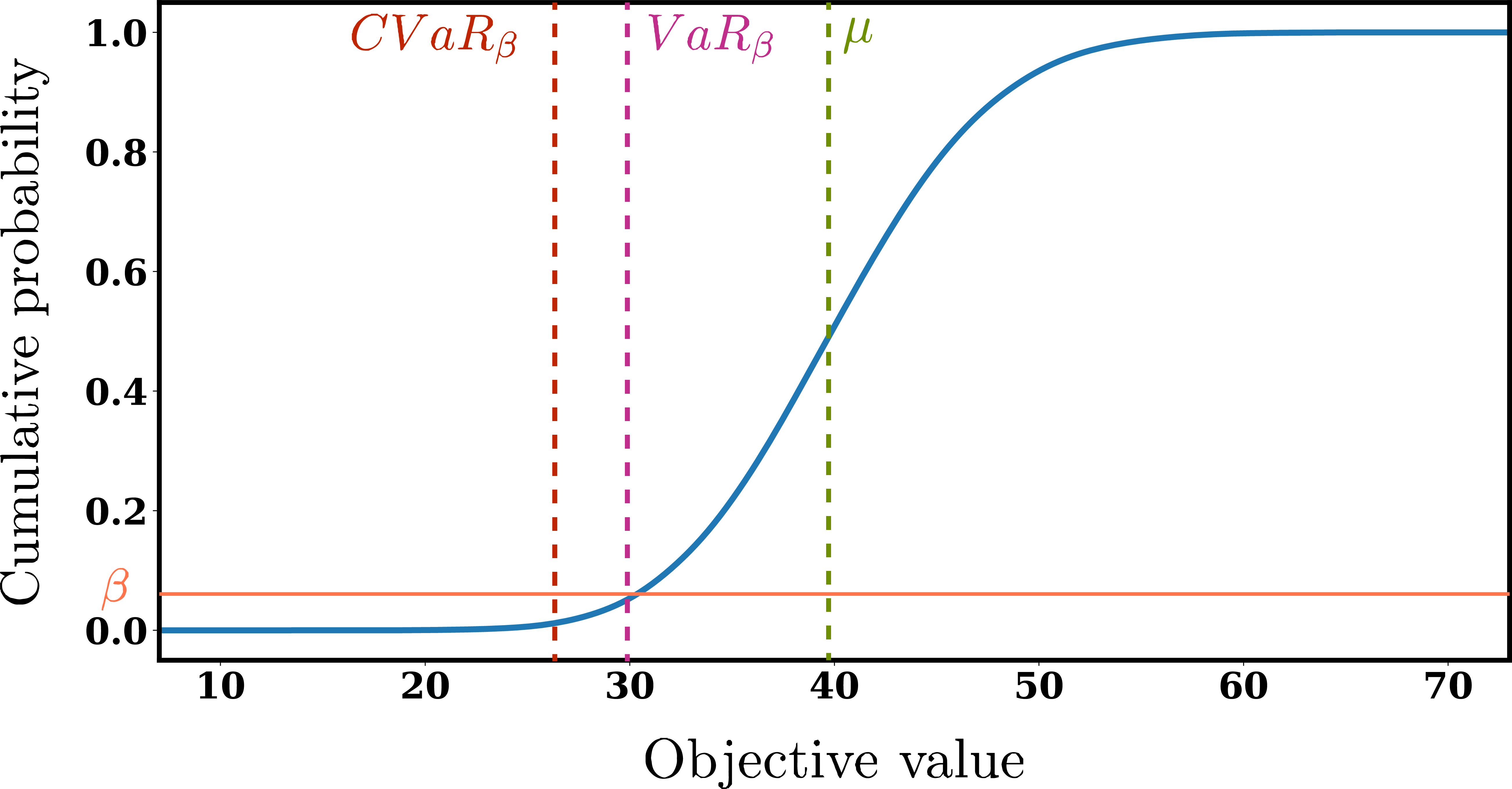}}\label{fig:cvarcdf}}%
    \caption{Description of the conditional value-at-risk, $CVaR_{\beta}$, and the value-at-risk, $VaR_{\beta}$, for a given probability level $\beta$, as well as the expected value, $\mu$ under a) the probability density function, $p_\pi(z^\phi)$, and b) the cumulative distribution function, $F_\pi(z^\phi)$.}
    \label{fig:CVAR}
\end{figure}

In the context of sequential decision-making problems, it is important to note that the CDF, and hence the VaR, is dependent on the policy, $\pi$. The following definitions are provided in terms of reward maximization, rather than loss (given the context of RL and MDPs). Specifically, the VaR is defined:
\begin{equation}\label{eq:VaR}
    \begin{aligned}
    z_\beta^\phi = \text{max} \{z^\phi: F_\pi(z^\phi) \leq \beta \} \quad \Longleftrightarrow \quad  z_\beta^\phi = F^{-1}_\pi(\beta)
    \end{aligned}
\end{equation}
It follows then that the CVaR is the expected value, $\mu_\beta^\phi\in \mathbb{R}$, of the random variable, $Z^\phi$, with a probability less than or equal to $\beta$ under $F_\pi(z^\phi)$:
\begin{equation}\label{eq:CVaR}
    \begin{aligned}
    \mu_\beta^\phi = z_\beta^\phi + \frac{1}{\beta} \int_{z^\phi<z_\beta^\phi} p_\pi(z^\phi)(z^\phi-z_\beta^\phi) dz^\phi 
    \end{aligned}
\end{equation}
Eq. \ref{eq:CVaR} expresses the CVaR (for a given probability $\beta$). This can be interpreted as the VaR minus the expected difference between the VaR and the returns, $z^\phi$, which are realized with probability $\leq \beta$, i.e. such that $F_\pi(z^\phi) \leq \beta$. This is further reinforced by Fig. \ref{fig:CVAR}, which provides a visualization of the VaR and CVaR. 

The optimization of the CVaR has advantage over the VaR in engineering applications because it provides more information about the performance of the policy within the tails of the distribution, $p_\pi(z^\phi)$. This is particularly beneficial if the distribution is characterised by heavy tails, which is often the case, given the possibility for rare events within the plant. Further, $Z_\beta^\phi$ possesses many undesirable functional properties that makes its optimization more difficult than $\mu_\beta^\phi$ \cite{rockafellar2002conditional, uryasev2000conditional}. Furthermore, the accuracy of the estimation of the $Z_\beta^\phi$ via samples, converges at a slower rate than $\mu_\beta^\phi$ \cite{hong2009simulating}. Therefore, in this work we favor the CVaR and present means to estimate from samples in the proposed framework. 
\subsubsection{Optimization of CVaR via sampling}
We seek means by which to gain approximations of Eqs. \ref{eq:VaR} and \ref{eq:CVaR} via sampling. Specifically, assuming one has $N$ realizations, $Z_{MC}^\phi = [z_1^\phi, \ldots, z_N^\phi]$, then Eq. \ref{eq:VaR} may be approximated via the $\floor*{\beta N}$\textsuperscript{th} order statistic, which we denote $\bar{z}_\beta^\phi$. The CVaR, $\mu_\beta^\phi$, may then be approximated via $\bar{\mu}_\beta^\phi$ as follows:
\begin{equation}
    \begin{aligned}\label{eq:samplesCVAR}
    \bar{\mu}_\beta^\phi = \bar{z}_\beta^\phi + \frac{1}{\beta N}\Bigg[\sum_{i=1}^N \text{min}(0, z_i^\phi - \bar{z}_\beta^\phi)\Bigg]
    \end{aligned}
\end{equation}
Here, we simply replace the integral term in Eq. \ref{eq:CVaR} via its sample average approximation. The \textit{minimum} operator enforces the bounds of integration detailed, such that only realizations less than the VaR are considered. The statistical properties associated with estimation of $\bar{\mu}_\beta^\phi$ have been well researched. Under some assumptions, the Monte Carlo estimate of the CVaR converges with increasing samples at the same rate as the sample average approximation \cite{hong2009simulating, pml2Book}. We direct the interested reader to the cited works for more information. 

How one incorporates this distributional perspective into the RL problem is highly flexible when applying stochastic search policy optimization. The CVaR may be either optimized as a sole objective or it may be enforced as a constraint subject to appropriate definition of $\beta$. In other words, one may formulate either of the two following optimization problems:
\begin{subequations}
\begin{equation}\label{eq:CVaRobj}
    \begin{aligned}
    \pi^*_\beta &= \argmax_{\pi} \bar{\mu}_\beta^\phi 
    \end{aligned}
\end{equation}
\begin{equation}\label{eq:CVaRcons}
    \begin{aligned}
    \pi^*_\beta &= \argmax_{\pi} \mathbb{E}_{\pi}\big[Z^\phi\big] \quad s.t.\quad \bar{\mu}_\beta^\phi \geq b
    \end{aligned}
\end{equation}
\end{subequations}
where $b\in \mathbb{R}$ is some minimum desired value. For the use of RL, we are dependent upon obtaining a closed form expression as an optimization objective. Optimization of Eq. \ref{eq:CVaRobj} is trivial. Whereas Eq. \ref{eq:CVaRcons} may be handled via a penalty method, or simply by discarding infeasible policies in the case of stochastic search algorithms. Please see \cite{uryasev2000conditional} for more information on CVaR optimization formulations. A general approach to optimization of either of the formulations (Eqs. \ref{eq:CVaRobj} and \ref{eq:CVaRcons}) is provided by Algorithm \ref{alg:generalSSPO}. The description is kept general to ensure clarity of the methodology. For further information on stochastic search optimization and the algorithm used in this work, PSO-SA, please see \ref{app:PSO-SA}.

\vspace{0.2cm}
\begin{algorithm}[H]
\SetAlgoLined
\caption{A general approach to distributional stochastic search policy optimization}
\vspace{0.1cm}
\justify
\textbf{Input}: Policy function (a neural network in this study), $\pi(\hat{\theta},\cdot)$, parameterized by $\hat{\theta}$; a number of samples to evaluate each candidate policy, $n_I$; sample approximation of objective function to optimize, $f_{SA}(\cdot)$; a general stochastic search optimization algorithm, $f_{SSO}(\cdot)$; population size, $P$; upper, $\theta_{UB}$, and lower, $\theta_{LB}$, bounds on the search space; population initialization method, $f_{init}(\hat{\theta}, P, \theta_{UB}, \theta_{LB})$; a number of optimization iterations, $K$; a memory buffer, $\mathcal{B}_{SSO}$; memory for metrics of optimal policy, $\mathcal{B}_{\pi^*} =\{J_{\pi^*}, \pi(\theta^*,\cdot)\}$\vspace{0.2cm}\\
\textbf{1.} Generate initial (neural network) population parameters, $\Theta^1 = f_{init}(\hat{\theta},P, \theta_{UB}, \theta_{LB})$, where $\Theta^1 = \{\theta_1, \ldots, \theta_P\}$\vspace{0.1cm};\\
\textbf{2.} \For{k = 1, \ldots, K}{\justify
\textbf{a.} Construct policy population, $\Pi^{k}= \{\pi(\theta_i,\cdot), \forall \theta_i \in \Theta^k\}$\vspace{0.1cm};\\
\textbf{b.} \For{each (neural network) policy $\pi_{c,i}\in \Pi^k$}{
                    \textbf{i.} Evaluate \textit{penalized return} of the policy, $Z^\phi$, of $\pi_{c,i}$ via Algorithm \ref{alg:rollout} for $n_I$ samples\vspace{0.1cm}\;
                    \textbf{ii.} Return distribution information of the policy, $\mathcal{B}_{\pi_{c,i}} = \{\mathcal{B}_{info}^{1},\ldots,\mathcal{B}_{info}^{n_I}\}$ from
                    
                    Algorithm \ref{alg:rollout}\vspace{0.1cm}\;
                    
                    \textbf{iii.} Assess sample approximate objective, $J_i = f_{SA}(\mathcal{B}_{\pi_{c,i}})$\vspace{0.1cm}\;
                    
                    \textbf{iv.} Collect information for policy $\pi_{c,i}$ and append $[J_i,\mathcal{B}_{\pi_{c,i}}]$ to $\mathcal{B}_{SSO}$\vspace{0.1cm}\;
                    
                    \textbf{v.} \textbf{if} $J_i > J_{\pi^*}$ \textbf{then} update best known policy $\mathcal{B}_{\pi^*} = \{J_i, \pi_{c,i}\}$\;}
\justify\textbf{c.} Generate new  parameters $\Theta^{k+1} = f_{SSO}(\mathcal{B}_{SSO})$, where $\Theta^{k+1} = \{\theta_{c,1}, \ldots, \theta_{c,P}\}$\vspace{0.1cm}\;}
\justify\textbf{Output:}  $\pi(\theta^*,\cdot)\in \mathcal{B}_{\pi^*}$ \vspace{0.1cm}
\label{alg:generalSSPO}
\end{algorithm} 

To provide demonstration of the ideas discussed, we now investigate the application of the methodology on a case study, which has been adopted from early work provided in \cite{cerda1997mixed}. We add plant uncertainties to the case study and construct a discrete-time simulation of the underlying plant. The case study is discussed further in the next Section. 

\section{Case Studies}\label{sec:CS}
\subsection{Problem definition}
We consider a multi-product plant where the conversion of raw material to product only requires one processing stage. \textit{We assume} there is an unlimited amount of raw material, resources, storage and wait time (of units) available to the scheduling element. Further, the plant is completely reactive to the scheduling decisions of the policy, $\pi$, although this assumption can be relaxed (with appropriate modification to the method stated here) if decision-making is formulated within an appropriate framework as shown in \cite{hubbs2020deep}. The scheduling element \textit{must decide} the sequencing of tasks (which correspond uniquely to client orders) on the equipment (units) available to the plant. The production scheduling environment is characterized by the following properties and requirements: 
\begin{enumerate}
    \item A given unit \textit{l} has a maximum batch size for a given task \textit{i}. Each task must be organized in campaigns (i.e. processed via multiple batches sequentially) and completed once  during the scheduling horizon. All batch sizes are predetermined, but there is uncertainty as to the processing time (this is specific to task and unit).
    \item The task should be processed before the delivery due date of the client, which is assumed to be an uncertain variable (the due date is approximately known at the start of the scheduling horizon, but is confirmed with the plant a number of periods before the order is required by the client).
    \item There are constraints on the viable sequencing of tasks within units (i.e. some tasks may not be processed before or after others in certain units).
    \item There is a sequence and unit dependent cleaning period required between operations, during which no operations should be scheduled in the given unit.
    \item Each task may be scheduled in a subset of the available units.
    \item Some units are not available from the beginning of the horizon and some tasks may not be processed for a fixed period from the start of the horizon (i.e. they have a fixed release time).
    \item Processing of a task in a unit must terminate before another task can be scheduled in that unit.
\end{enumerate}

The objective is to minimize the makespan and the tardiness of task (order) completion. Once all the tasks have been successfully processed according to the operational rules defined, then the decision making problem can be terminated. The problem is modelled as an MDP with a discrete-time formulation. The original work \cite{cerda1997mixed} utilized a continuous-time formulation. Further discussion is provided later in the text on this topic. Full details of the MDP construction, uncertain state space model and model data used in this work is provided by \ref{app:model}.

\subsection{Benchmark}
The benchmark for the following experiments as presented in Section \ref{sec:CSexp} is provided by both offline and online implementations of the continuous time, mixed integer linear programming (MILP) model first detailed in \cite{cerda1997mixed}.

To ensure that the solutions from the two time transcriptions (i.e. the continuous time formulation of the benchmark and the discrete-time formulation of the methodology proposed) are comparable, the data which defines the sequence dependent cleaning times, task-unit dependent processing times and release times are redefined from the original study to ensure that their greatest common factor is equal to the length of a time interval in the discrete-time transcription. 

Online implementation was dependent upon updating the model data by receiving feedback from the state of the plant at a given discrete time index. 
Given that there are uncertainties in the plant dynamics, the MILP model utilizes the expected values of the uncertain data. Similarly, in RL expected values of uncertain variables are maintained in the state representation, until the uncertainty is realized at which point the state is updated appropriately \cite{hubbs2020deep}. Please see \cite{cerda1997mixed} and \ref{app:data} for more information on the model and data used for the following experiments, respectively. All MILP results reported were generated via the Gurobi v9.1.2 solver together with the Pyomo v6.0.1 modelling framework. The proposed method utilized the PyTorch v1.9.0 python package  and Anaconda v4.10.3.
\subsection{Experiments}\label{sec:CSexp}
\subsubsection{Problem instances and sizes}
In the following, we present the formulation of a number of different experiments across two different problem instances. The first problem instance is defined by a small problem size. Specifically, we are concerned with the sequencing of 8 client orders (tasks) on four different units. This corresponds to 304 binary decision variables and 25 continuous decision variables within the benchmark continuous-time MILP formulation. The second problem instance investigates the ability of the framework to handle larger problems with 15 orders and 4 units. In this case, the MILP formulation consists of 990 binary decision variables and 46 continuous decision variables.

\subsubsection{Study designs and assessment of performance}
Both of the problem instances are investigated thoroughly. We demonstrate the ability of the method to handle: a) uncertainty in processing times; b) uncertainty in the due date; and, c) the presence of finite release times. These three considerations, a)-c), are used as factors to construct a full factorial design of experiments. Each factor has two levels, either it is present in the underlying problem, or it is not. As a result, the majority of analysis focuses on the investigation of the method's ability to optimize in expectation. This is captured by experiments E1-E8 in Table \ref{table:exp_conds}.

A number of the experimental conditions were used to demonstrate the ability of the method to optimize for the CVaR of the distribution also. This was investigated exclusively within problem instance 1, as detailed by experiments D3-D8 in Table \ref{table:exp_conds}. 

In all experiments, the processing time uncertainty is defined via a uniform probability density function (see Eq. \ref{eq:uniformpt}) and the due date uncertainty is described by a Poisson distribution (see Eq. \ref{eq:poissondue_date}). 
\begin{table}[h]
  \caption{Table of experimental conditions investigated. Details of the exact descriptions of uncertain variables are provided by \ref{app:model}.}
  \label{table:exp_conds}
  \small
  \centering
  \begin{tabular}{cccc}
    \toprule
    \multicolumn{4}{c}{Optimizing in Expectation (Eq. \ref{eq:expectationopt})} \\
    \midrule
    \multicolumn{4}{c}{Problem Instance 1 \& 2}\\
    \midrule
    Reference & Processing time uncertainty & Due date uncertainty & Finite release times \\
    \midrule
    E1 & \xmark & \xmark & \xmark \\
    E2 & \xmark & \xmark & \checkmark\\
    E3 & \xmark & \checkmark & \xmark \\
    E4 & \xmark & \checkmark & \checkmark \\
    E5 & \checkmark & \xmark & \xmark \\
    E6 & \checkmark & \xmark & \checkmark \\
    E7 & \checkmark & \checkmark & \xmark \\
    E8 & \checkmark & \checkmark & \checkmark \\
    \toprule
    \multicolumn{4}{c}{Optimizing for Conditional Value at Risk (CVaR) (Eq. \ref{eq:CVaRobj})} \\
    \midrule
    \multicolumn{4}{c}{Problem Instance 1}\\
    \midrule
    Reference & Processing time uncertainty & Due date uncertainty & Finite release times \\
    \midrule
    D3 & \xmark & \checkmark & \xmark \\
    D4 & \xmark & \checkmark & \checkmark \\
    D5 & \checkmark & \xmark & \xmark \\
    D6 & \checkmark & \xmark & \checkmark \\
    D7 & \checkmark & \checkmark & \xmark \\
    D8 & \checkmark & \checkmark & \checkmark\\
    \bottomrule
  \end{tabular}
\end{table}

Further to the experiments proposed in Table \ref{table:exp_conds}, the proposed and benchmark methods are compared with respect to online and offline computational burden. The robustness of the proposed RL method to misspecification of the plant uncertainties is also investigated. The results are detailed in Section \ref{sec:timecost} and \ref{sec:robustRL}, respectively.

The performance indicators used to evaluate the respective methods include the expected performance of the scheduling policy, $\mu_z = \mathbb{E}_{\pi}\big[Z\big]$, the standard deviation of the performance, $\sigma_Z=\Sigma_{\pi}\big[Z\big]$ and the conditional-value-at-risk. Specifically, we use a version of the CVaR, $\bar{\mu}_\beta$, which considers the non-penalised returns\footnote{This is assessed via appropriate modification to Eq. \ref{eq:samplesCVAR}, such that we consider the sum of rewards, rather than the return under the penalty function.} with $\beta =0.2$ (i.e. $\bar{\mu}_\beta$ represents the expected value of the worst case policy performance observed with probability less than or equal to 0.2). Finally, we utilize a statistically robust approximation to the probability of constraint satisfaction, $F_{LB}$. Formal details of evaluation of $F_{LB}$ are provided by \ref{app:constraintssamples}.  
In the following section, the results for the different experiments are presented for the method proposed and benchmarked relative to a continuous-time MILP formulation. 

\section{Results and Discussion}\label{sec:R&D}
In this section, we turn our attention to analysing the policy training process and ultimate performance of the framework proposed. In all cases, results were generated by the hybrid PSO-SA stochastic search algorithm with space reduction. In \textit{learning} (or stochastic search), all candidate policies were evaluated over $n_I=50$ samples (when the plant investigated was subject to uncertainty, if deterministic $n_I=1$), with a population size of $P=60$ and maximum optimization iterations of $K=150$. The structure of the network utilized for policy paramaterization is detailed by \ref{app:PSO-SA}. 

\subsection{Policy training}

\begin{figure}[h]
\centering
  \subfloat[\centering Expected training profile \vspace{0.1cm}]{{\includegraphics[scale=0.13]{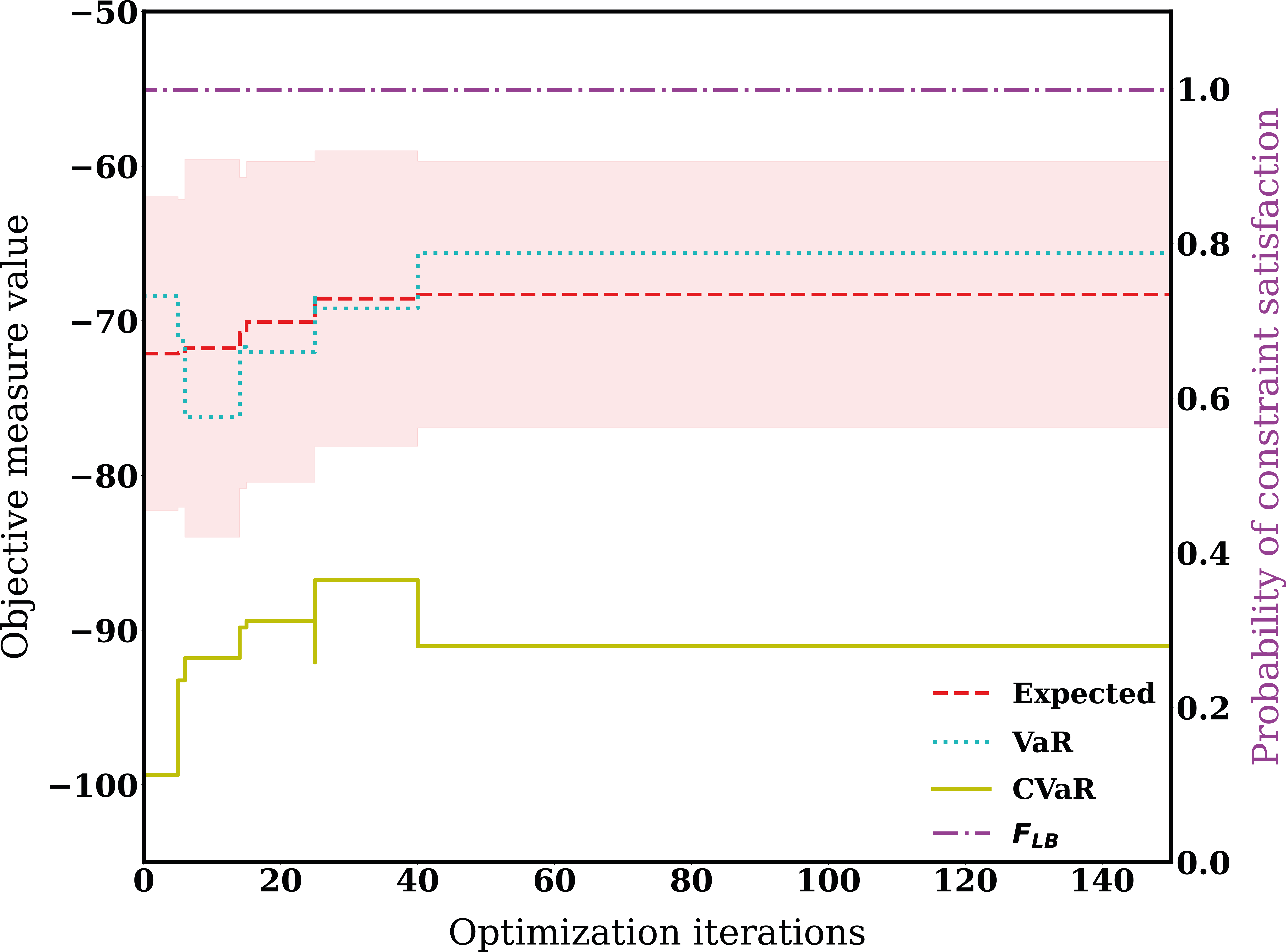}}\label{fig:RL_mean_train}}%
  \quad
\centering
  \subfloat[\centering Distributional training profile\vspace{0.1cm}]{{\includegraphics[scale=0.13]{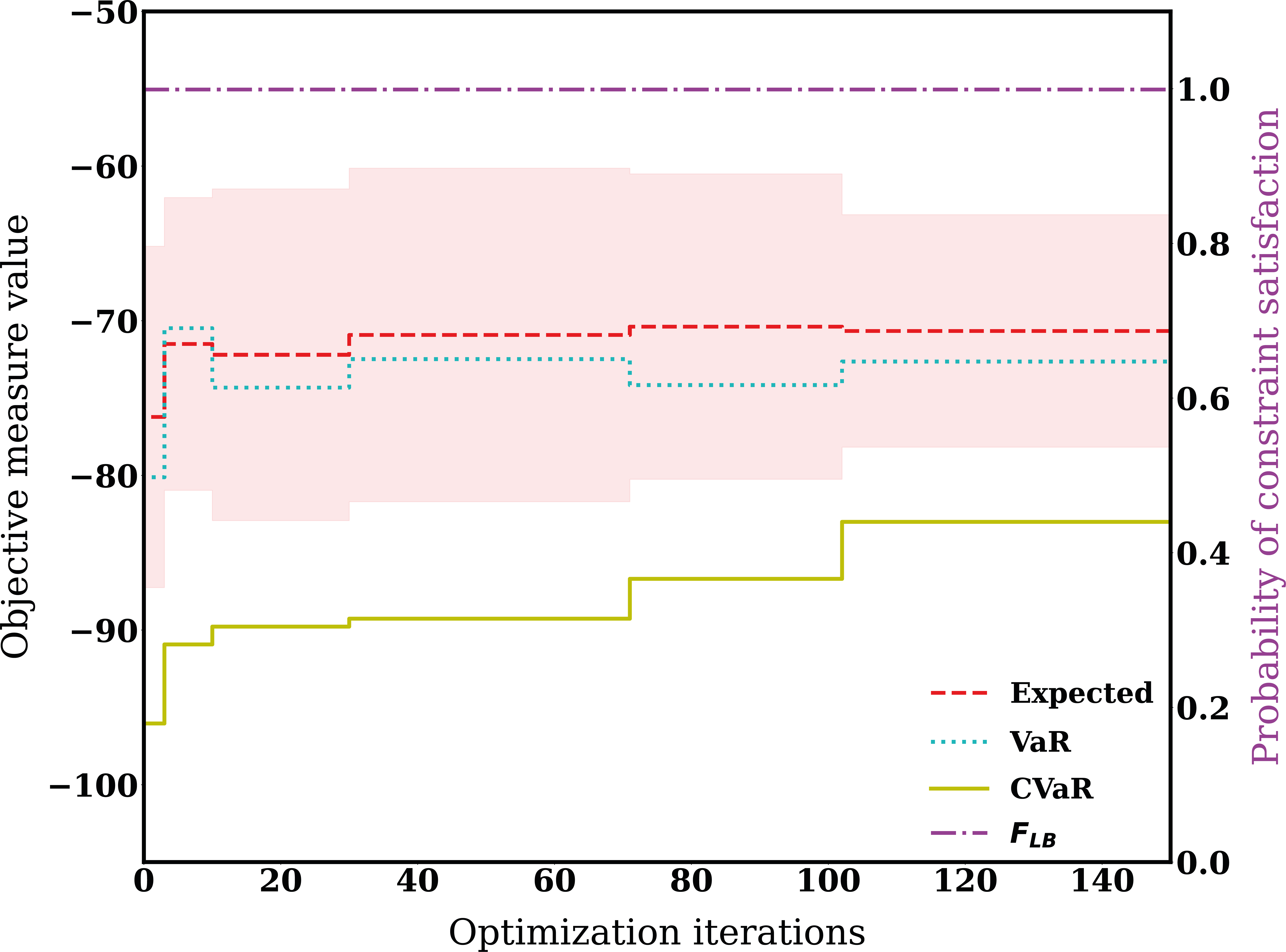}}\label{fig:RL_dist_train}}%
\caption{The training profile of the RL agent on experiment E8, problem instance 1. Metrics of the best known policy are tracked as the population is iterated. Plot a) shows the mean, standard deviation (shaded region around the expected profile), value-at-risk ($\beta=0.2$), the corresponding conditional-value-at-risk and probability of constraint satisfaction, $F_{LB}$, for formulation Eq. \ref{eq:expectationopt}. Plot b) displays the same information for formulation Eq. \ref{eq:CVaRobj}.} 
\label{fig:RL_training}
\end{figure}

Demonstration of the training profiles for experiment E8, problem instance 1 are provided by Fig. \ref{fig:RL_training}. Fig. \ref{fig:RL_mean_train} details the formulation provided by Eq. \ref{eq:expectationopt} and Fig. \ref{fig:RL_dist_train} details Eq. \ref{eq:CVaRobj} (i.e. the expected and CVaR objective, respectively). 
The plots detail how the methodology steadily makes progress with respect to the measure of the objective posed by the respective problem formulation. The VaR represents the maximum objective performance observed with probability less than or equal to 0.2. Hence, the similarity between the VaR and the expected performance, as expressed by Fig. \ref{fig:RL_training}, indicates that the returns are not well described by a Gaussian distribution \cite{wilcox2003applying}. This is an approximation required by many distributional RL algorithms, but not stochastic search optimization approaches as used in this work. Optimization iterations proceed after population initialization. Therefore, the performance of the initial best known policy is dependent upon the initialization. Thereafter, each iteration consists of evaluating each candidate policy in the population over $n_I=50$ samples, such that 150 iterations is equivalent to obtaining 450,000 simulations of the uncertain model. The computational time cost of this is discussed in Section \ref{sec:timecost}. All current best known policies identified throughout training satisfy the constraints imposed on the problem, indicating the efficacy of framework proposed.

\subsection{Problem instance 1}\label{sec:P1}
In this section, the results of investigations for experiments corresponding to problem instance 1 are presented. In this problem instance there are 8 customer orders and 4 units available to the production plant. Firstly, we present results confirming the ability of the method to identify an optimal scheduling solution for a deterministic plant, which corresponds to the generation of an offline production schedule. We then turn our attention to scheduling of the plant subject to uncertainties. 

\subsubsection{Optimization of the deterministic plant (offline production scheduling)}

There are two experiments that investigate the optimality of the method proposed in generation of an offline production schedule: E1 and E2 (see Table \ref{table:exp_conds}). Investigation E1 defines a deterministic plant without finite release times, whereas E2 includes finite release times. 

\begin{figure}[h]
\centering
  \subfloat[\centering Experiment E1: RL solution\vspace{0.2cm}]{{\includegraphics[scale=0.29]{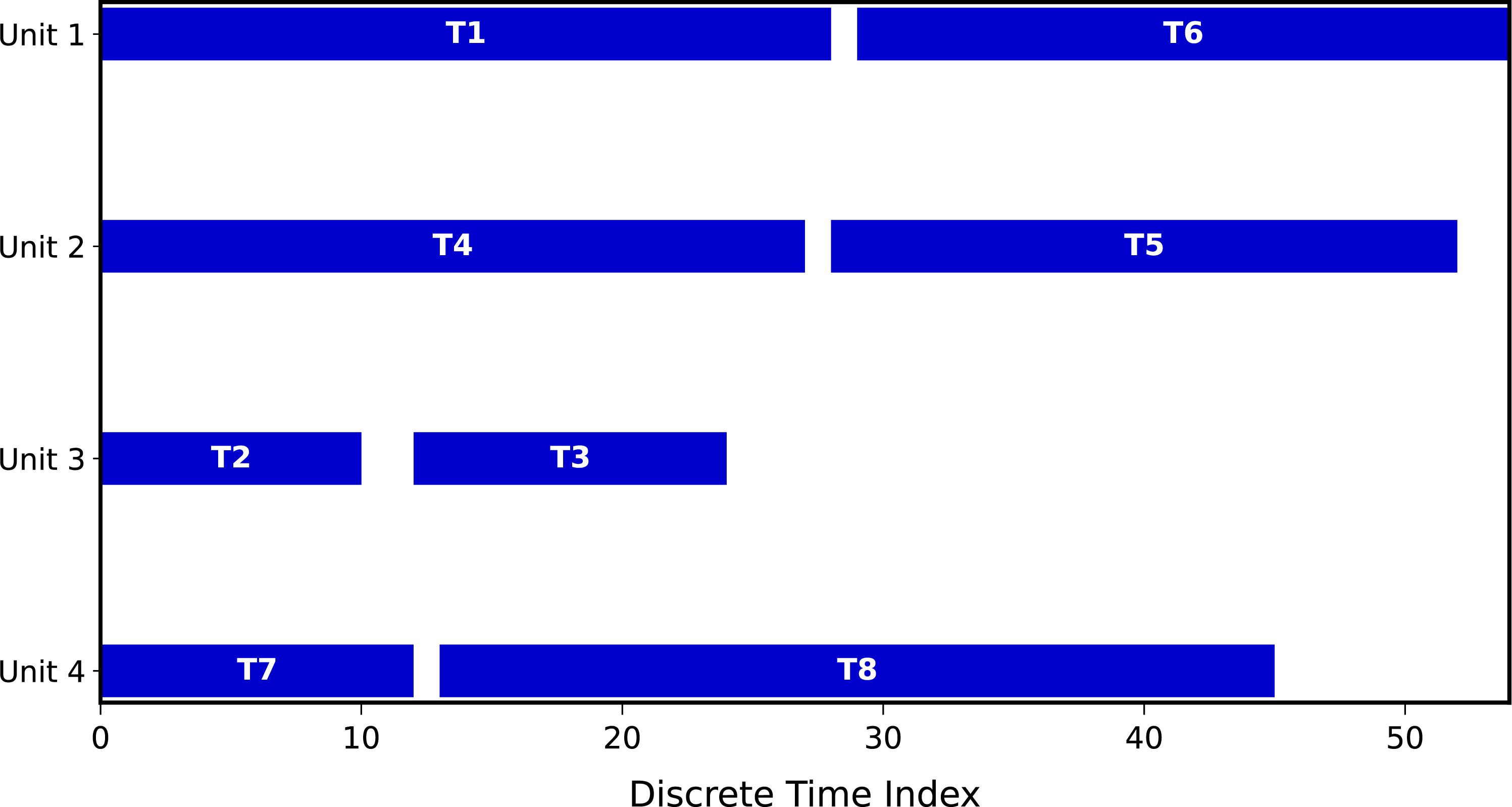}}\label{fig:RL_P1_E1}}%
  \quad
\centering
  \subfloat[\centering Experiment E1: MILP solution\vspace{0.2cm}]{{\includegraphics[scale=0.29]{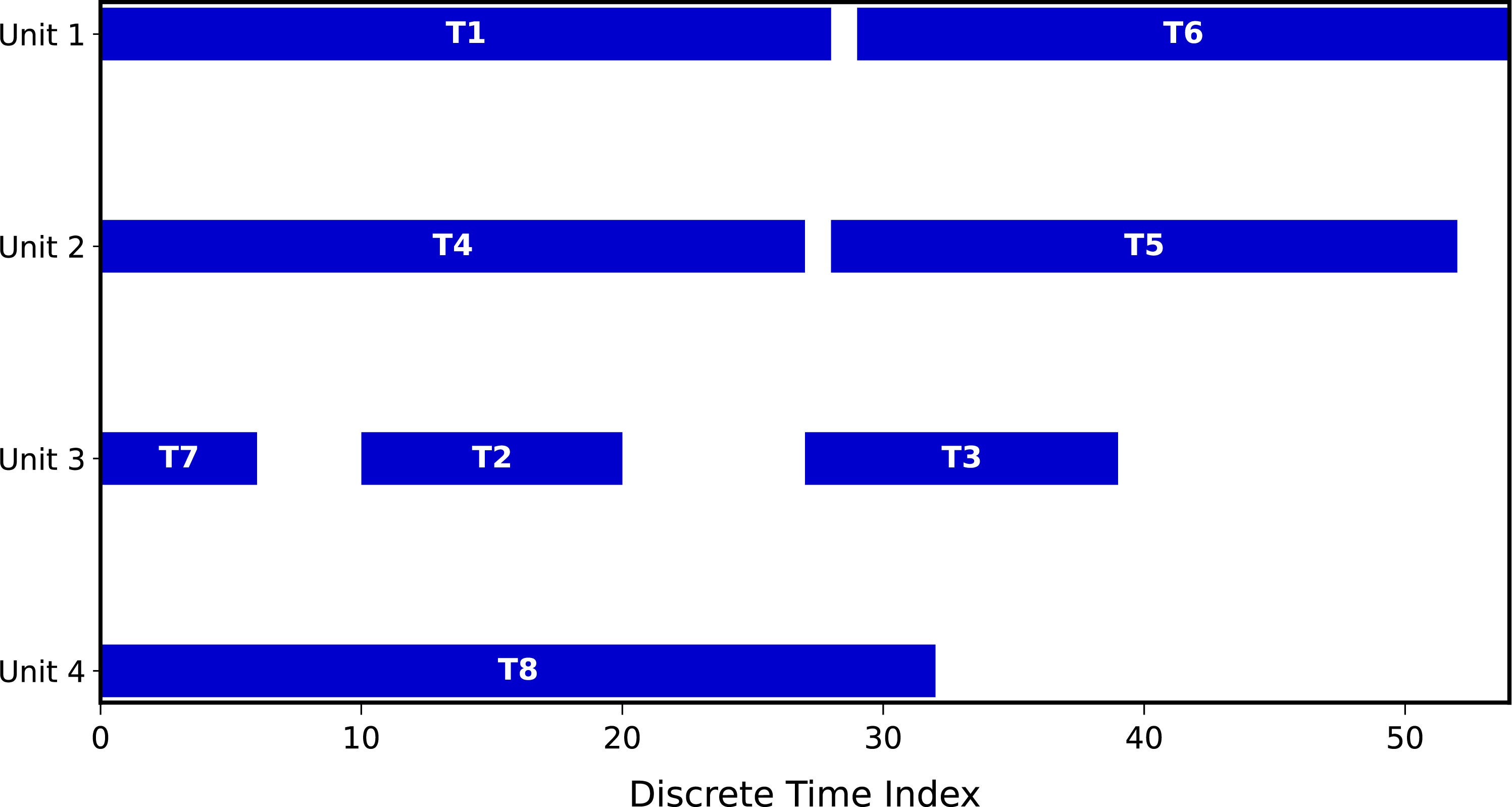}}\label{fig:MILP_P1_E1}}%
  \newline
  \newline
  \centering
  \subfloat[\centering Experiment E2: RL solution]{{\includegraphics[scale=0.29]{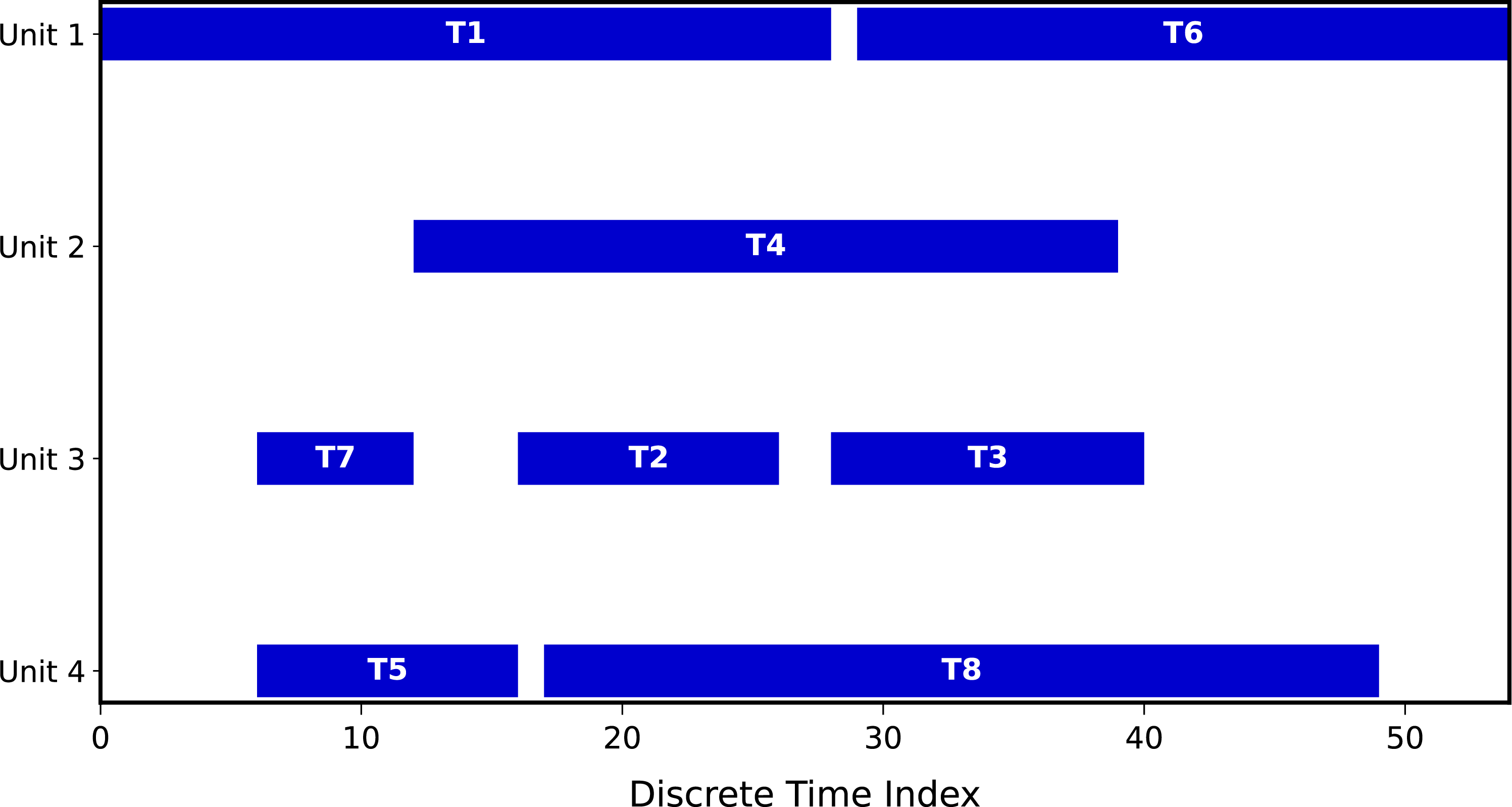}}\label{fig:RL_P1_E2}}%
  \quad
\centering
  \subfloat[\centering Experiment E2: MILP solution]{{\includegraphics[scale=0.29]{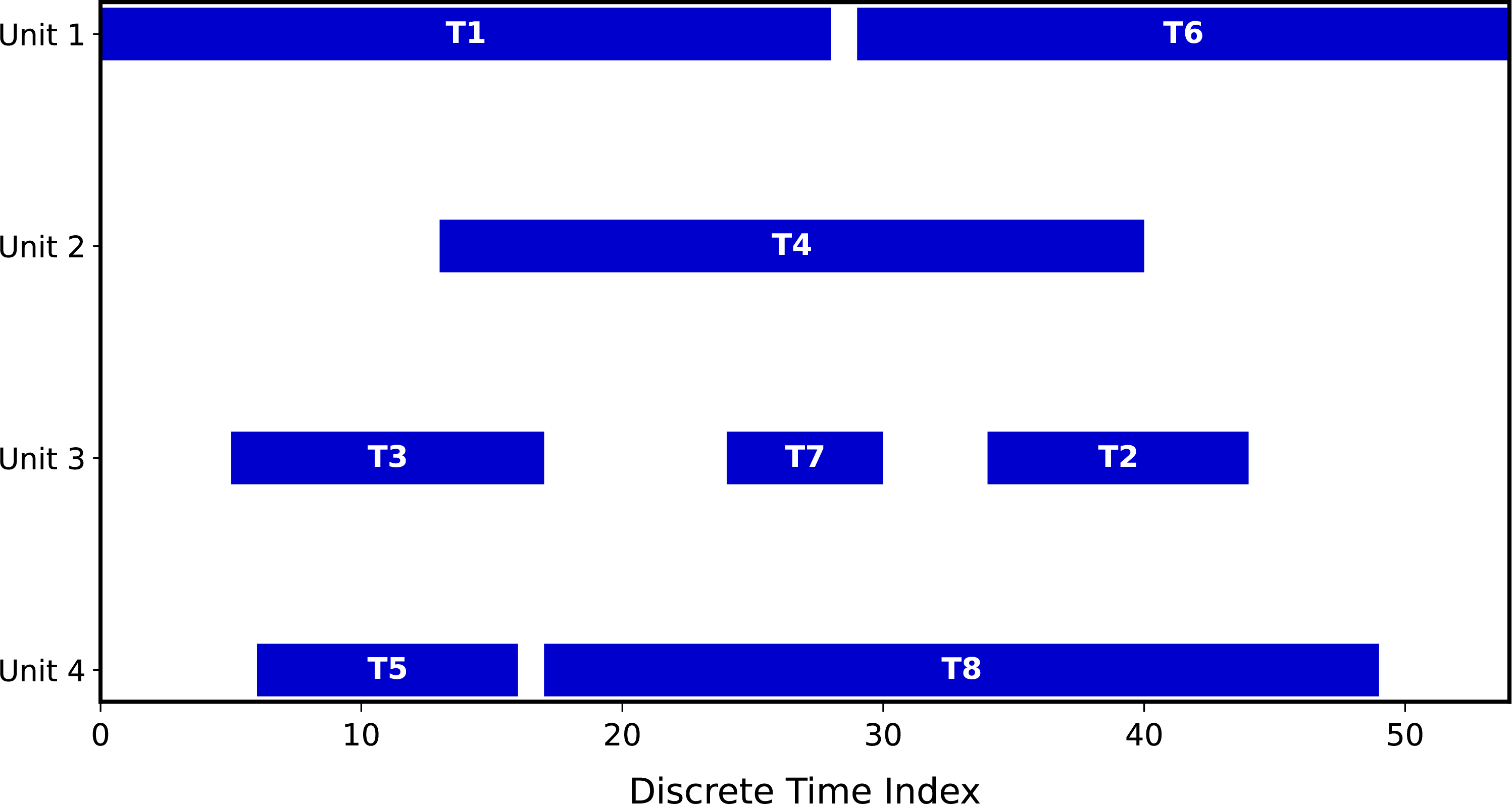}}\label{fig:MILP_P1_E2}}%
\caption{Investigating the offline schedule generated for the deterministic plant. The results for experiment E1, problem instance 1 generated by the a) RL and b) MILP methods. The results for experiment E2, problem instance 1 generated by the c) RL and d) MILP methods. The label T\textit{i} details the scheduling of task \textit{i} in a given unit.} 
\label{fig:E1n2P1}
\end{figure}

Fig. \ref{fig:RL_P1_E1} and \ref{fig:MILP_P1_E1} provides a comparative plot of the schedule generated via the policy identified via the method proposed and the MILP formulation for experiment E1, respectively. 
As detailed from Table \ref{table:expres}, the policy obtained from the method proposed achieves the same objective value as the benchmark MILP method, despite the slight difference in the structure of the solutions identified. The difference in the RL solution arises in the sequencing of tasks in unit 3 and 4, but does not affect the overall quality of schedule identified. Importantly, this result highlights the ability of the method proposed to account for e.g. sequence dependent cleaning times, sequencing constraints, and requirements to complete production in campaigns in the appropriate units as imposed in this case study

Next, we turn our attention to the handling of finite release time within the framework provided. This is probed by experiment E2. Again, a comparative plot of the relative production schedules identified by the MILP formulation and proposed method is provided by Fig. \ref{fig:RL_P1_E2} and \ref{fig:MILP_P1_E2}. Despite a slight difference in the structure of the solution, arising from the sequencing of tasks in unit 3; as detailed from Table \ref{table:expres}, the policy obtained from the method proposed achieves the same objective value as the benchmark MILP method on experiment E2. This further supports the capacity of the framework to handle the constraints imposed on the problem. In the following section, we explore the ability of the framework to handle plant uncertainties.
\subsubsection{Optimization of the uncertain plant (reactive production scheduling)}\label{sec:onlineP1}
In this section, we turn our attention to demonstrating the ability of the framework to handle plant uncertainty. Specifically, we incrementally add elements of plant uncertainty to the problem via investigation of experiments E3-E8. Again, the MILP formulation is used to benchmark the work. For both approaches, policies were validated by obtaining 500 Monte Carlo (MC) simulations of the policy under the plant dynamics. This enables us to accurately estimate measures of the distribution of return, $p_{\pi}(z)$. 
\begin{table}[h]
  \caption{Table of results for the proposed method from investigation of experimental conditions detailed by Table \ref{table:exp_conds} for Problem Instance 1. The policies synthesised were optimized under the objective provided by Eq. \ref{eq:expectationopt}.}
  \label{table:expres}
  \small
  \centering
  \begin{tabular}{c|c|cccc|c|cccc}
    \toprule
    Reference & Method & $\mu_Z$  & $\sigma_Z$ & $\bar{\mu}_\beta$ & $F_{LB}$ & Method & $\mu_Z$  & $\sigma_Z$ & $\bar{\mu}_\beta$ & $F_{LB}$ \\
    \midrule
    E1 & \multirow{8}{*}{Proposed} & $\textbf{-62.0}$ & 0.0 & -62.0 & 1.0  & \multirow{8}{*}{MILP} & -62.0 & 0.0 & -62.0 & 1.0\\
    E2 &  & $\textbf{-65.0}$ & 0.0 & -65.0 & 1.0 &  & -65.0 & 0.0 & -65.0 & 1.0\\
    E3 &  & \textbf{-61.9} & 4.4 & -72.5 & 1.0 &  & -63.3  & 4.4 &  -72.2  & 1.0\\
    E4 &  & \textbf{-66.0} & 4.9 & -75.6 & 1.0 &  &  -66.3  & 4.9 & -76.5 & 1.0\\
    E5 &  & \textbf{-66.8} & 8.7 & -86.0 & 1.0 &  & -70.1  & 9.6  & -90.2  & 1.0\\
    E6 &  & -73.8 & 10.7 & -97.5 & 1.0 &  & \textbf{-73.6}  & 10.3  & -94.0 & 1.0\\
    E7 &  & \textbf{-67.4} & 10.9 & -86.8 & 1.0 &  & -71.6 & 11.3  & -93.5 & 1.0\\
    E8 &  & -75.3 & 11.5 & -101.13 & 0.99 &  & \textbf{-75.1} & 11.7  & -97.7 & 1.0\\
    \bottomrule
  \end{tabular}
\end{table}

Table \ref{table:expres} presents the results for the method proposed and the MILP benchmark for experiments E3-8. Here, the proposed method aims to optimize $\mu_Z$ (the expected performance), whereas the MILP formulation assumes the plant is deterministic, and only accounts for plant uncertainty via state-feedback. In 4 out of 6 experiments (E3, E4, E5 and E7), the RL method outperforms the MILP approach. The most significant of these is experiment E7, where the proposed approach (-67.4) outperforms the MILP (-71.6) by 5.8\% in the objective. The MILP approach only marginally outperforms the proposed method (by $\leq 0.2$\%) in 2 of 6 of the experiments. This is observed in the results of experiment E6 and E8.

Further, in all but one of the experiments, constraints are respected absolutely across all 500 MC simulations by both the methods. In experiment E8, however, the method proposed violated the constraints in what equates to 1 out of the 500 MC simulations. This is primarily because in the optimization procedure associated with the method candidate policies are evaluated over 50 MC simulations. This means that a realization of uncertainty, which caused the policy to violate the constraint expressed by the penalty function in validation, was not observed in the \textit{learning phase} (see \ref{app:constraintset} for more information regarding the specific constraint). This can be mitigated practically by increasing the number of samples, $n_I$, that a policy is evaluated for, at the cost of increased computation.

Most of the analysis so far has concentrated on the expected performance, $\mu_Z$, of the policy and the probability of constraint satisfaction, $F_{LB}$. Generally, all standard deviation, $\sigma_Z$, of $p_{\pi}(z)$ across the experiments are similar for both the RL approach and the MILP approach. There is some discrepancy between the CVaR, $\bar{\mu}_\beta$, of the RL and MILP which considers the worst 20\% of returns associated with $p_{\pi}(z)$, however, this should be interpreted with care given that there is no justification for either method's formulation to be better than the other. However, this can be incentivized in the RL approach and this is discussed in the subsequent section.

\subsubsection{Optimizing risk-sensitive measures}
In this section, we present the results from investigation of experiments D3-8. The difference between these experiments and E3-8, is that now we are interested in optimizing for a different measure of the distribution, $p_{\pi}(z)$. In E3-8, the objective was optimization in expectation, $\mu_Z$. In D3-8, the objective is optimization for the expected cost of the worst 20\% of the penalised returns, i.e. Eq. \ref{eq:CVaRobj}, with $\beta=0.2$. Again, the optimization procedure associated with the proposed method utilizes a sample size of $n_I=50$. All policies were then evaluated for 500 MCs.
\begin{table}[h]
  \caption{Results for distributional RL corresponding to investigation of experimental conditions detailed by Table \ref{table:exp_conds} for Problem Instance 1. Results that are emboldened detail those policies that show improved CVaR, $\bar{\mu}_\beta$, for $\beta =0.2$, over the MILP approach (as detailed in Table \ref{table:expres}).}
  \label{table:distresults}
  \small
  \centering
  \begin{tabular}{c|ccccc}
  \toprule
  Method & Reference & $\mu_Z$  & $\sigma_Z$ & $\bar{\mu}_\beta$ & $F_{LB}$ \\
  \midrule
  \multirow{5}{*}{Proposed} & D3 & -62.1 & 4.1 & \textbf{-69.8} & 0.99\\
    &D4 & -65.5 & 4.9 & \textbf{-75.5} & 0.99\\
    &D5 &-67.2 & 8.2 & \textbf{-83.9} & 1.0 \\
    &D6 &-74.9 & 11.2 & -99.0 & 1.0\\
    &D7 & -67.9 & 9.7 & \textbf{-87.1} & 0.99 \\
    &D8 &-73.2 & 11.6 & \textbf{-96.7} & 1.0 \\
    \bottomrule
  \end{tabular}
\end{table}

Table \ref{table:distresults} details the results of the investigations D3-8. As before, the results corresponding to $\bar{\mu}_\beta$ quantify the CVaR of the sum of rewards, $Z$, i.e. not including penalty for violation of constraints. The results that are emboldened show improvements in the CVaR over the MILP results as detailed in Table \ref{table:expres}. To reiterate, the only difference here is that the MILP uses the expected values of data in its formulation, whereas the distributional RL approach (the proposed method) accounts for uncertainty in the data and optimizes for the expected value of the worst 20\% of penalized returns. This leads to improvements in the CVaR in 5 out of 6 of all experiments (D3-8), with an average improvement of 2.29\% in the expected value of the worst 20\% of the returns. This figure should be interpreted with care, given that there is likely to be some statistical inaccuracy from the sampling process. However, the weight of the result supports that there is an improvement in performance gained from the RL formulation.

The expected performance, $\mu_Z$, of the policy identified for experiments D3-8 is also competitive with the MILP formulation. However, this is not promised to hold generally, given improvements in the expected performance are not incentivized within the formulation (i.e. Eq. \ref{eq:CVaRobj}, although this result is also noted in \cite{sarin2014minimizing}). To balance both objectives the formulation provided by Eq. \ref{eq:CVaRcons} could be investigated, if it is of interest to a given production setting.  
\begin{figure}[h]
\centering
  \subfloat[\centering Histogram of objective performance  \vspace{0.1cm}]{{\includegraphics[scale=0.13]{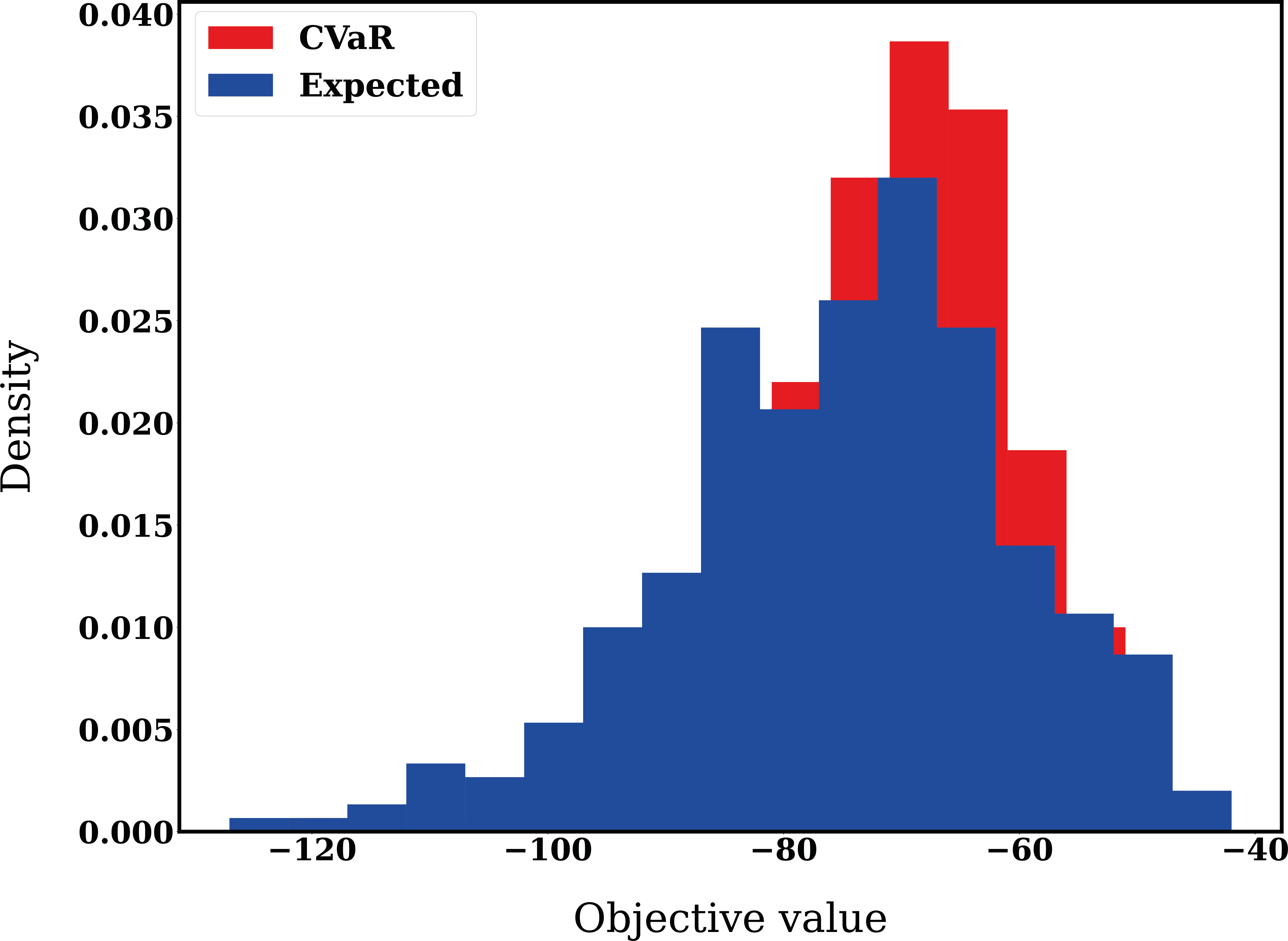}}\label{fig:RL_epdf}}%
  \quad
\centering
  \subfloat[\centering Empirical cumulative distribution function \vspace{0.1cm}]{{\includegraphics[scale=0.13]{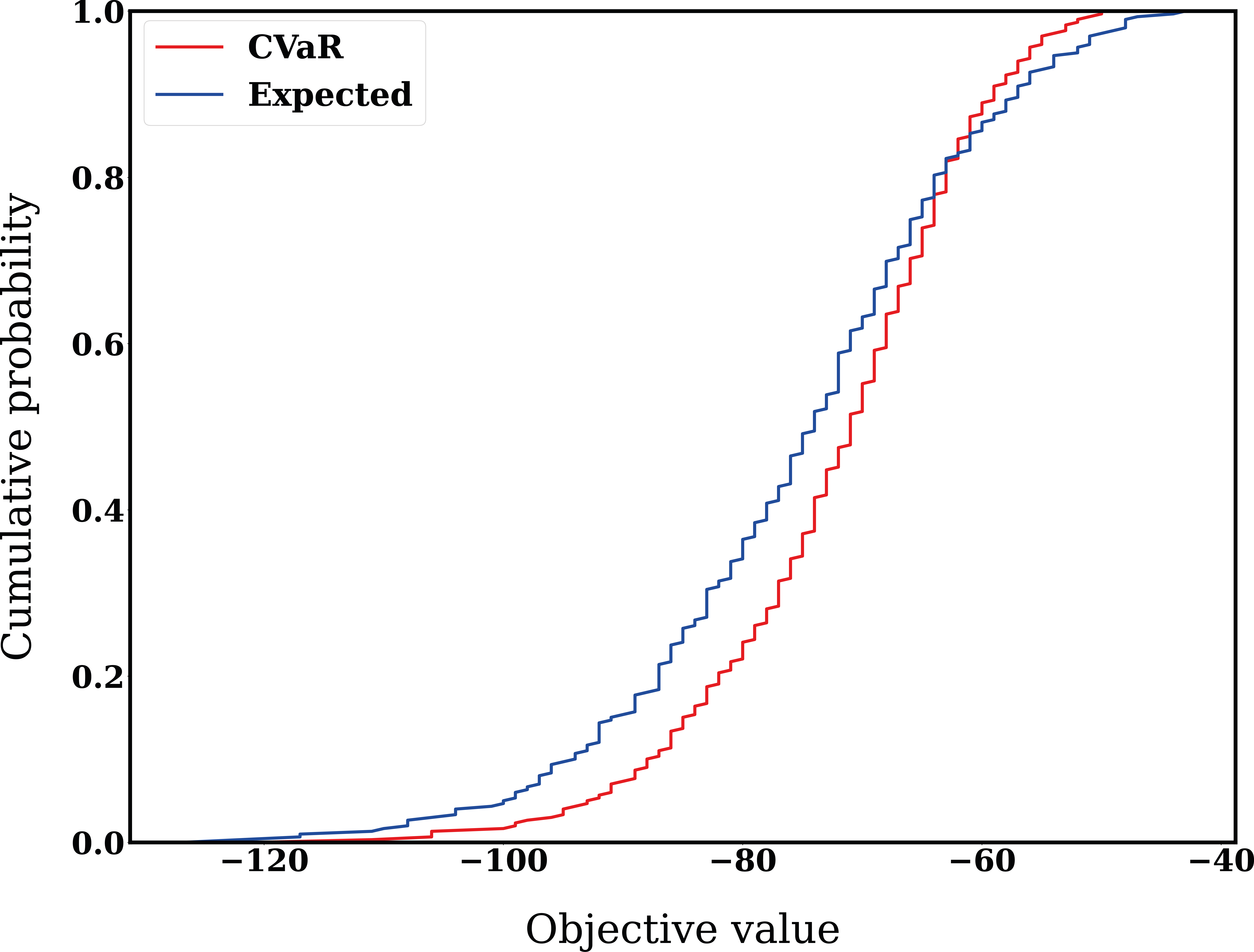}}\label{fig:RL_ecdf}}%
\caption{The distributions of returns observed in validation of the RL policy obtained from optimizing expectation (i.e. E8) and conditional value-at-risk (i.e. D8) within the same production environment. Plot a) shows a histogram of the returns, and b) shows a plot of the empirical cumulative distribution function associated with each policy.} 
\label{fig:DistRLimprov}
\end{figure}
\FloatBarrier

Similar observations can be made with respect to the probability of constraint satisfaction as in Section \ref{sec:onlineP1}. In 3 of the 6 experiments, the constraint imposed in the penalty function is violated in approximately 1 of the 500 MC simulations. Again, this could be mitigated by increasing the number of samples, $n_I$, in policy evaluation during the optimization procedure. However, even at the current setting of $n_I=50$, the constraints are respected with high probability, and this was deemed sufficient for this work.

Fig. \ref{fig:DistRLimprov} expresses the distribution of returns under the RL policies optimizing expectation and CVaR, as obtained in experiment E8 and D8, respectively. Both policies are learned in production environments defined by the same dynamics and uncertainties. Their quantitative metrics can be found in Table \ref{table:expres} and \ref{table:distresults}, respectively. Particularly, Fig. \ref{fig:RL_ecdf} highlights how the CVaR policy observes improved performance in the tail of the distribution over the policy optimizing for expectation (i.e. the CVaR plot is shifted along the x-axis to the right). Although, the policy does not observe as good a best-case performance, this eloquently highlights the utility of distributional formulations to identify risk-sensitive policies. 

\subsection{Problem instance 2}\label{sec:P2}

\subsubsection{Optimization of the deterministic plant (offline production scheduling)}
Here, the results of investigation of experiments E1 and E2 are presented for problem instance 2. The purpose of the investigation is to determine whether the method is able to scale with larger problem instances efficiently and retain a zero optimality-gap as in Section \ref{sec:P1}. Both experiments consider the plant to be deterministic, with E2 additionally considering the presence of finite release times. 

\begin{figure}[h]
\centering
  \subfloat[\centering RL solution\vspace{0.2cm}]{{\includegraphics[scale=0.29]{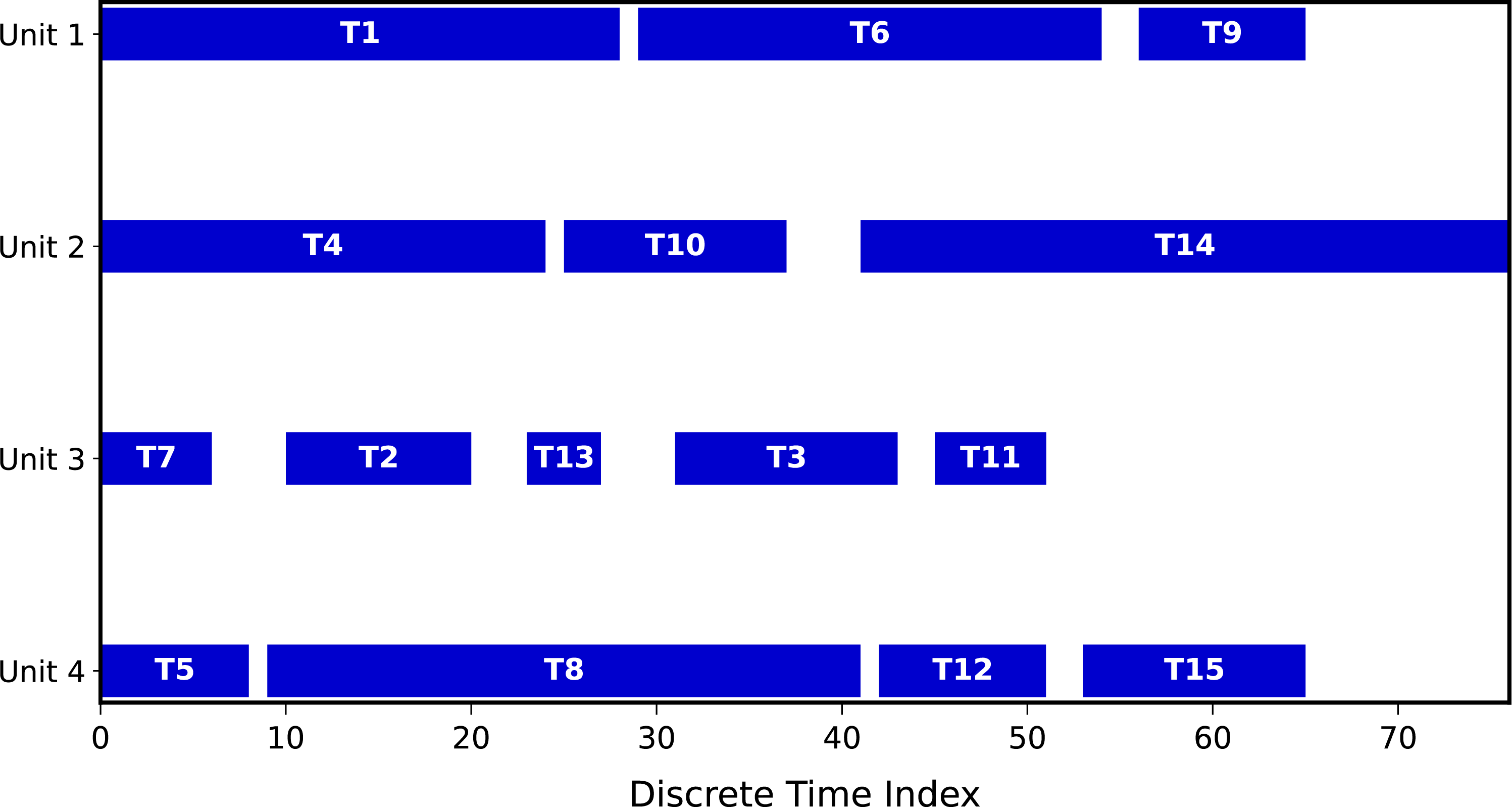}}\label{fig:RL_P2_E1}}%
  \quad
\centering
  \subfloat[\centering MILP solution\vspace{0.2cm}]{{\includegraphics[scale=0.29]{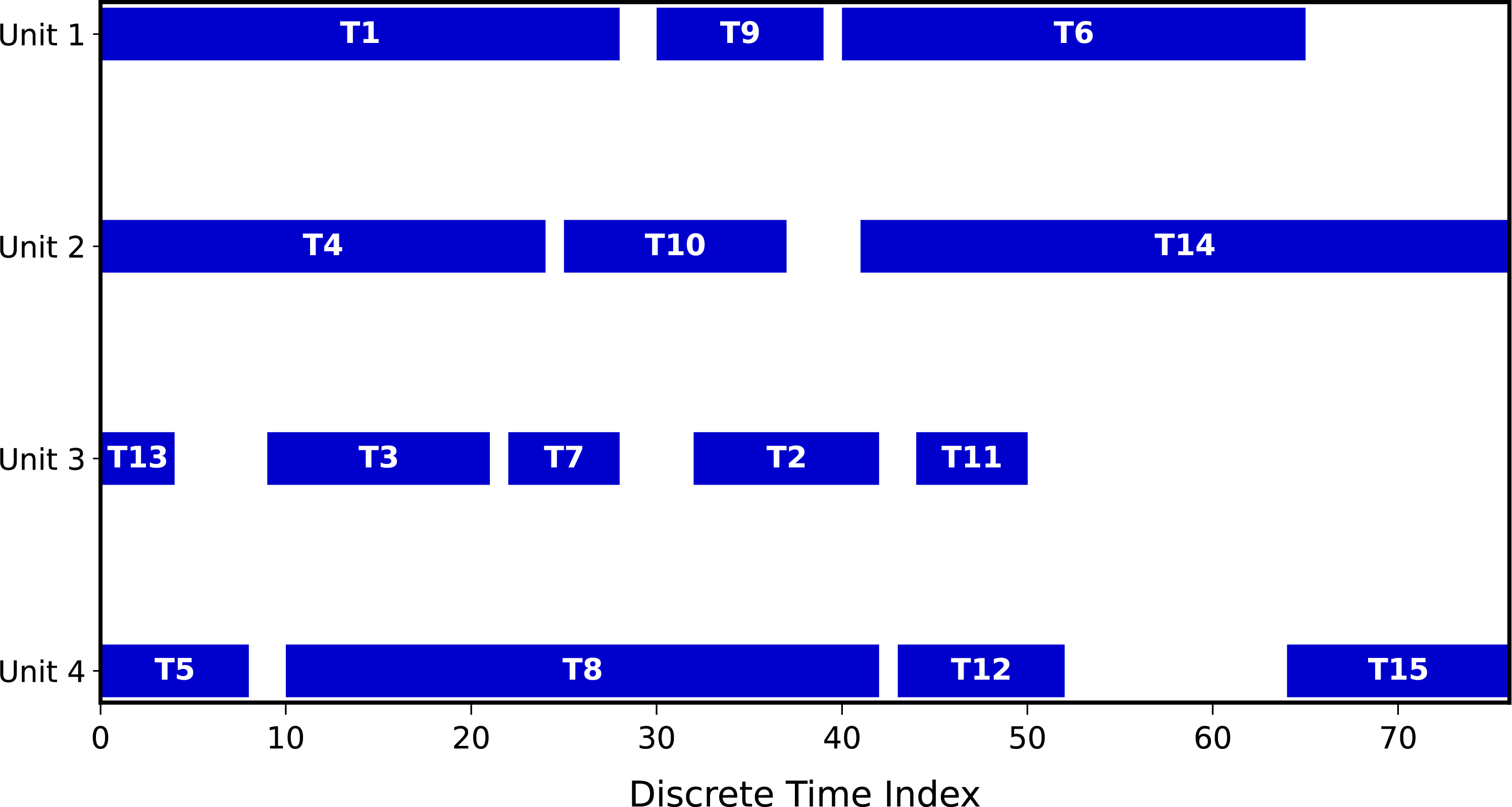}}\label{fig:MILP_P2_E1}}%
  \newline
  \centering
  \subfloat[\centering RL solution\vspace{0.2cm}]{{\includegraphics[scale=0.29]{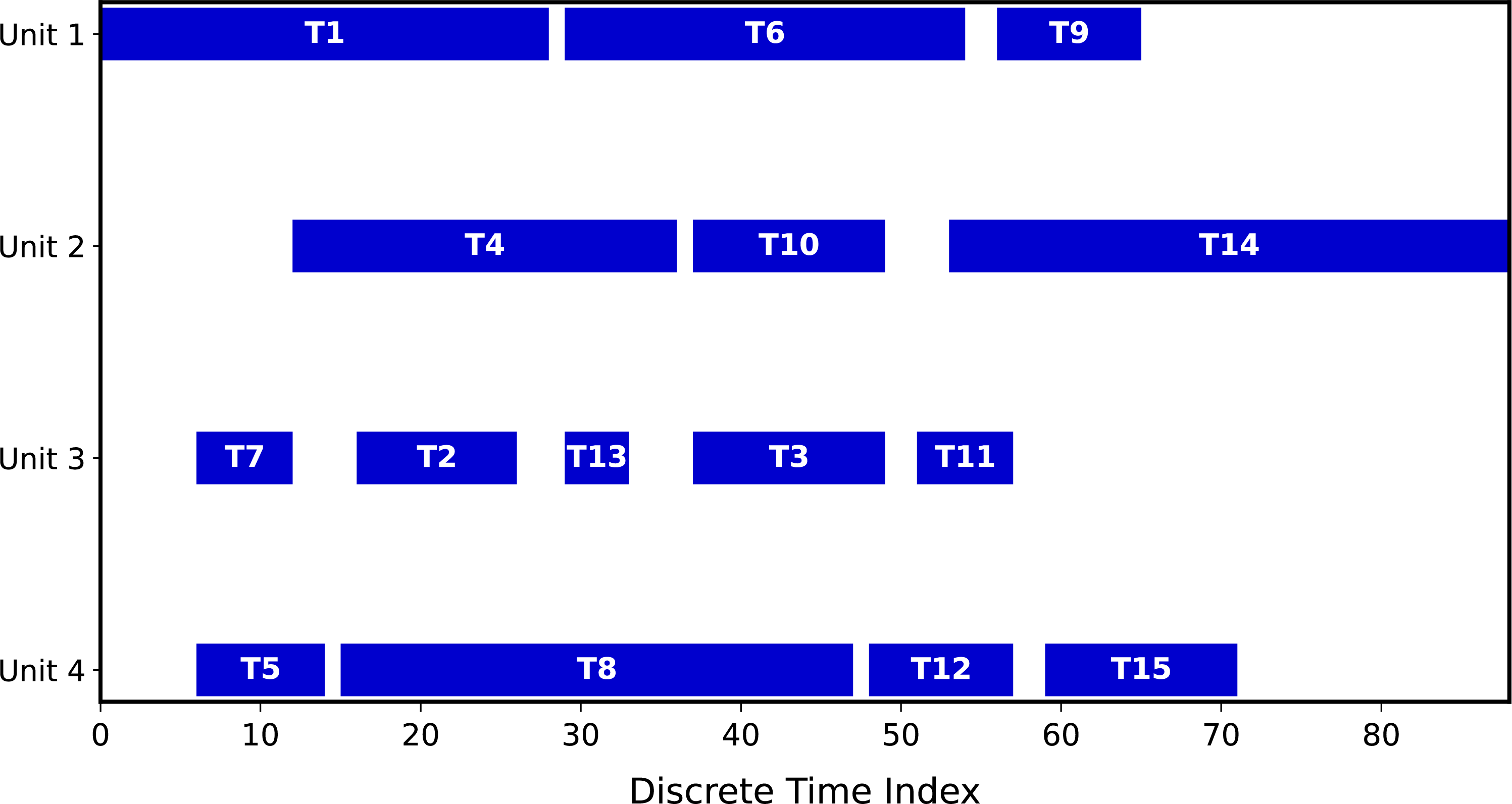}}\label{fig:RL_P2_E2}}%
  \quad
\centering
  \subfloat[\centering MILP solution\vspace{0.2cm}]{{\includegraphics[scale=0.29]{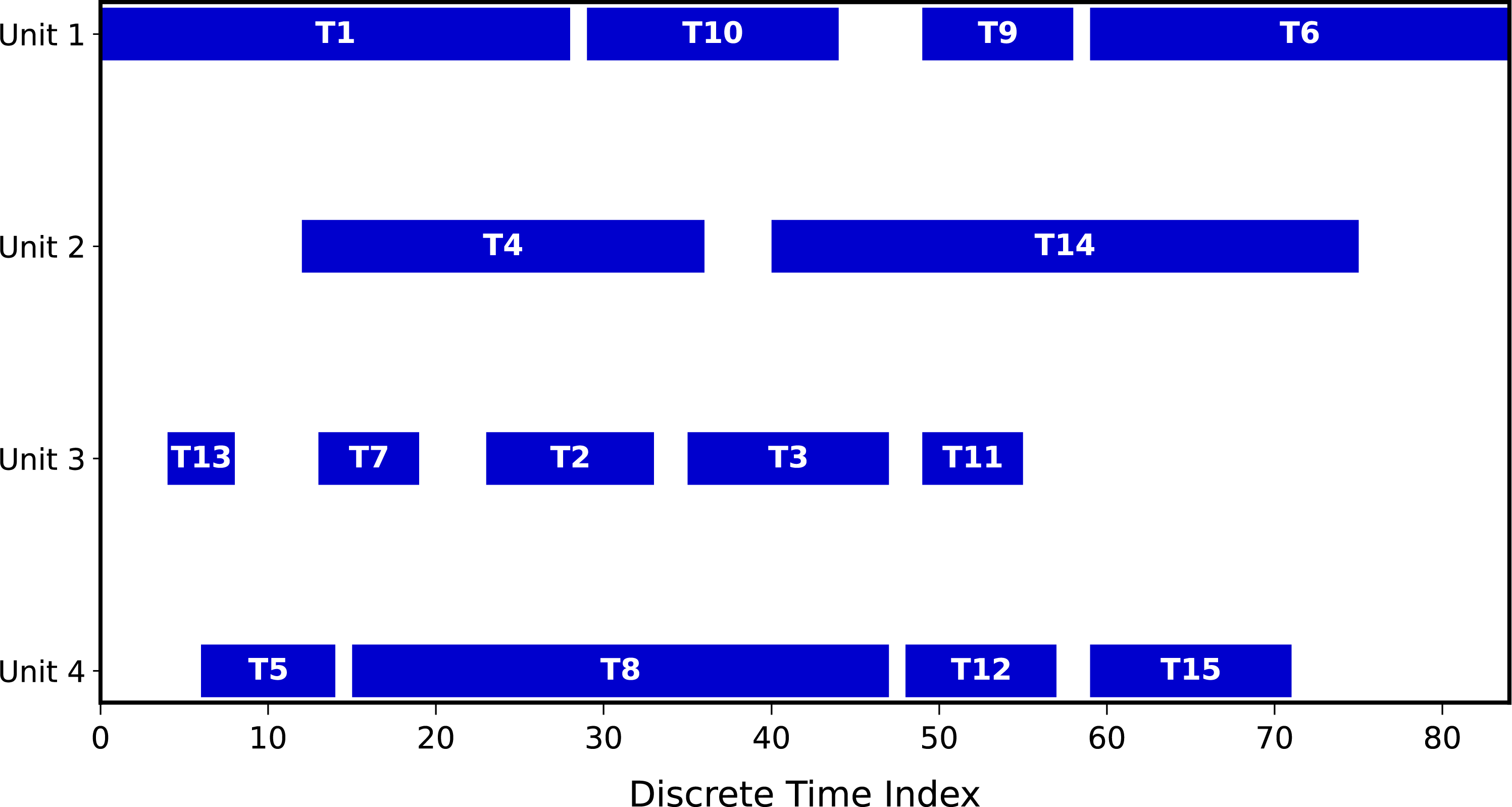}}\label{fig:MILP_P2_E2}}%
\caption{Investigating the offline schedule generated for the deterministic plant. The results for experiment E1, problem instance 2 generated by the a) RL and b) MILP methods. The results for experiment E2, problem instance 2 generated by the c) RL and d) MILP methods. The label T\textit{i} details the scheduling of task \textit{i} in a given unit.}
\label{fig:E12P2}
\end{figure}

Fig. \ref{fig:RL_P2_E1} and \ref{fig:MILP_P2_E1} presents the comparative results of the schedule identified for experiment E1 of the RL and MILP approach. There are clear differences between the schedules. These arise in the sequencing of task 6 and 9 in unit 1, and in the sequencing of unit 3. However, for this case the RL approach is able to retain a zero optimality-gap. This provides promise for the method proposed in scaling to larger problem sizes. This is reinforced by the value of the objective as presented in Table \ref{table:expresPI2}, where both methods achieve an objective score of -107. 
Fig. \ref{fig:RL_P2_E2} and \ref{fig:MILP_P2_E2} presents comparative results of the schedule generated for experiment E2. Here, we see that again there is slight difference between the schedule generated for the RL relative to the MILP. However in this case, there is an optimality gap of 4.4\% in the RL schedule generated. This is confirmed by Table \ref{table:expresPI2}, which details the RL method achieved an objective score of -143.0, whereas the MILP method achieved a score of -137.0. The difference between the schedules, which causes the optimality-gap, is observed in the scheduling of task 10 in unit 2 (in the case of RL). This leads to a slight increase in the overall makespan of production. This could be attributed to the complex and non-smooth mapping between the state and optimal control, which makes model structure selection a difficult task when using parametric models, such as neural networks. An alternate hypothesis exists in the parameter space, rather than the model space. The mapping between policy parameters and objective is again likely to be highly non-smooth, which could pose difficulties for stochastic search optimization. It should be noted, however, that the latter proposition should hold true for the experiment without finite release times (E1), but in this case an optimality gap is not observed. One could therefore consider devising algorithms to: a) enable use of function approximators more suited to expressing non-smooth relationships between the state and optimal control e.g. decision trees; b) automate the identification of the appropriate parametric model structure; or c) improve on the existing hybrid stochastic search algorithm used in this work.

Despite the presence of the optimality gap in experiment E2, in both experiments, the framework is able to efficiently handle the constraints imposed on the plant. In the following section, we consider the addition of process uncertainties via investigation of experiments E3-8. 

\subsubsection{Optimization of the uncertain plant (reactive production scheduling)}

From Table \ref{table:expresPI2}, it is clear that generally the MILP outperforms the RL formulation proposed. On average, the performance gap is 1.75\%. This can be decomposed, such that if we consider those experiments with finite release time present, the average gap is 2.73\%; for those experiments that did not include finite release times, the average gap is just 0.77\%. This further supports the argument that the optimality gap is dependent upon the underlying complexity of the optimal control mapping (i.e. the optimal scheduling policy becomes more difficult to identify when finite release times are included). 
\begin{table}[h]
  \caption{Table of results for the proposed method from investigation of experimental conditions detailed by Table \ref{table:exp_conds} for Problem Instance 2. The policies synthesised were optimized under the objective provided by Eq. \ref{eq:expectationopt}.}
  \label{table:expresPI2}
  \small
  \centering
  \begin{tabular}{c|c|ccc|c|ccc}
    \toprule
    Reference & Method & $\mu_Z$  & $\sigma_Z$ & $F_{LB}$ & Method & $\mu_Z$  & $\sigma_Z$ & $F_{LB}$ \\
    \midrule
    E1 & \multirow{8}{*}{Proposed} & \textbf{-107.0} & 0.0 & 1.0  & \multirow{8}{*}{MILP} & -107.0 & 0.0 & 1.0\\
    E2 &  & {-143.0} & 0.0 & 1.0 &  & \textbf{-137.0} & 0.0 &  1.0\\
    E3 &  & {-113.3} & 8.5  & 0.98 &  & \textbf{-108.4}  & 7.15 & 1.0\\
    E4 &  & {-150.5} & 10.2 & 0.97 &  &  \textbf{-142.7}  & 11.43 &  1.0\\
    E5 &  & \textbf{-119.4} & 16.4 & 1.0 &  & -123.4  & 17.1 &  1.0\\
    E6 &  & -167.1 & 21.9 & 0.98 &  & \textbf{-166.6}  & 20.6  & 1.0\\
    E7 &  & {-127.2} & 19.6 & 0.95 &  & \textbf{-125.9} & 19.5  & 1.0\\
    E8 &  & -171.6 & 21.7 & 0.99 &  & \textbf{-167.5} & 25.1  & 1.0\\
    \bottomrule
  \end{tabular}
\end{table}

It is also of note that the probability of constraint satisfaction decreases in this case study relative to problem instance 1, across experiments E3-8 (see Table \ref{table:expres}). However, these changes are generally small, with the largest change equivalent to a 4\% decrease. This is likely due to realizations of uncertainty unseen in the training having greater impact in validation, given the increased complexity of the decision making problem. However, as previously stated, this could be overcome by increasing $n_I$ in training, at the cost of added computational complexity. 

\subsection{Computational time cost in policy identification and decision-making}\label{sec:timecost}

In this section, we are interested in the comparative time intensity of the RL approach proposed relative to the MILP approach. Having identified a model of the plant and the associated uncertainties, we are interested in how long it takes to identify: a) a control policy, $\pi$, (this is conducted offline) and b) a control decision, $\mathbf{u}_t$, conditional to observation of the plant state, $\mathbf{x}_t$, at discrete time index, $t$ (this is conducted online). The code was parallelized to minimize the time complexity of computation. In practice, the computational metrics reported in the following could be bettered considerably, with respect to policy identification, if the appropriate hardware were available.

In the case of a), in RL we consider the amount of time to conduct stochastic search optimization associated with the proposed method when the sample size $n_I=50$; whereas, the MILP incurs no time cost in identification of the policy, given the optimization formulation is essentially the policy, $\pi$. In b), we consider the time taken: for RL to conduct a forward pass of the neural network parameterization of the control policy; and, to solve an MILP problem. Table \ref{table:time_table} details the respective results for problem instance 1 and 2.
\begin{table}[h]
    \centering
    \small
    \caption{Normalized times for a) offline identification of a control policy, $\pi$, and b) identification of online scheduling decisions for problem instances 1 and 2.}
    \begin{tabular}{c|c|ccc}
    \toprule
        & \multicolumn{3}{c}{Time for identification}\\
        \midrule
        Object of identification & Problem instance & MILP & RL )\\
        \midrule
        \multirow{2}{*}{Control decision (online)} &1 & 1 & 0.0067\\
        & 2 & 1 & 0.002 \\
        \midrule
        \multirow{2}{*}{Control policy (offline)} & 1 & 0 & 1 \\
         & 2 & 0 & 1 \\
        \bottomrule
    \end{tabular}
    \label{table:time_table}
\end{table}

From Table \ref{table:time_table}, it is clear that the RL method demands a much greater investment in the computational synthesis of a scheduling policy, $\pi$, offline. Generally, policy synthesis required between 1-2 hours for problem instance 1 and 2 under the sample size, $n_I$, detailed by this work. The differences in time for offline policy identification between problems 1 and 2 arises due to an increased number of orders (leading to a longer makespan in problem instance 2). However, the work required online to update the control schedule, given a realization of plant uncertainty, is far lower in the case of RL than in the MILP formulation. For problem instance 1, the MILP is characterized by 304 binary decision variables and 25 continuous decision variables. In this case, inference of a control decision is 150 times cheaper via the RL method proposed. In problem instance 2, the MILP formulation consists of 990 binary decision variables and 46 continuous decision variables. For a problem of this size, the RL approach is 500 times cheaper than the MILP. The online computational benefits of RL are likely to be considerable for larger case studies, which is of significant practical benefit to the scheduling operation. 

\subsection{The effects of inaccurate estimation of plant uncertainties}\label{sec:robustRL}
To further consider the practicalities of RL, in this section, we consider the effects of inaccurately estimating the real plant uncertainties for the policy synthesised via experiment E8, problem instance 1. Specifically, we assume that in the offline model of the plant dynamics and plant uncertainties are as previous. While in the following experiments M1-8, we assume that actual plant uncertainties are different from that of the offline model. The processing time uncertainty has the same form (a discrete uniform probability distribution), but with the upper and lower bounds misspecified by an additive constant, $k_{pt}\in \mathbb{Z}_+$. Similarly, the rate of the Poisson distribution descriptive of the due date uncertainty is misspecified. Here, however, the misspecification is treated probabilistically, such that the rate is either unperturbed or misspecified by a constant, $k_{dd}\in \mathbb{Z}_+$ (this is either added or subtracted), with uniform probability. See \ref{app:uncertaintymisspec} for further details. 

In the respective experiments M1-8, $k_{pt}$ and $k_{dd}$ are defined to investigate the effects of different degrees of plant-model mismatch. In all cases, the policy identified from experiment E8 in problem instance 1 was rolled out in a plant where the true uncertainties are misspecified for 500 MC simulations. The experiments and their associated results are detailed by Table \ref{table:exp_misspec}.
\begin{table}[h]
  \caption{Table of experimental conditions investigated. In each experiment, we take the trained policy from experimental condition E8, problem instance 1 and evaluate its performance in a plant defined by different uncertainties. The degree of misspecification increases with experiment number. }
  \label{table:exp_misspec}
  \small
  \centering
  \begin{tabular}{ccccccc}
    \toprule
    Reference & Processing time, $k_{pt}$ & Due date, $k_{dd}$ & $\mu_Z$ & $\sigma_Z$ & $\bar{\mu}_\beta^\phi$ & $F_{LB}$ \\
    \midrule
    Model (E8) & 0 & 0 & -75.3 & 11.5 & -101.1 & 0.99\\
    M1 & 0 &  1& -74.8 & 12.3 & -101.3 & 0.98 \\
    M2 & 0 &  2 & -73.8 & 11.8 & -99.9 & 0.99\\
    M3 & 1 &   0& -76.0 & 13.3 & -106.8 & 0.95\\
    M4 & 1 &  1& -77.3 & 13.2 & -104.4 & 0.95\\
    M5 & 1 &  2 & -78.0 & 14.8 & -105.6 & 0.95\\
    M6 & 2 &  0& -83.9 & 17.6 & -121.7 & 0.78\\
    M7 & 2&  1& -81.5 & 17.5 & -116.2 & 0.82\\
    M8 & 2&  2& -83.4 & 18.7 & -118.5 & 0.82\\
    \bottomrule
   \end{tabular}
\end{table}

From Table \ref{table:exp_misspec}, it is clear that misspecification of plant uncertainties generally impacts the performance of the policy identified. The misspecifications identified are relatively mild in the case of due date uncertainty, however, they are more pronounced in the case of processing times. This is reflected in the results of the experiments, where little variation is observed in the policy performance when $k_{pt} < 2$. For example, the maximum deviation in terms of expected performance, $\mu_z$, when this condition is not imposed, is observed in experiment M5 where the corresponding decrease in $\mu_z$ from the performance observed in the offline model (experiment E8) is just 3.6\%. Similarly, the probability of constraint satisfaction, $F_{LB}$, remains high (slightly decreases to 0.95). Notably in experiments M1 and M2, the policy performance increases. As the estimated values for due dates are maintained in the state (fed to the policy), the misspecification is unlikely to induce a significantly different state distribution than the model and hence the result should be interpreted with caution (i.e. it likely arises from the statistical inaccuracies of sampling).

The effects on policy performance become notable when the processing time misspecification, $k_{pt}=2$. In all experiments M6-8, where this condition is imposed, the decrease in expected performance, $\mu_Z$, is in the region of 10\%. Additionally, the magnitude of the CVaR for $\beta=0.2$ (i.e. the expected value of the worst 20\% of policy performances) decreases by 15-20\%, and this is mirrored further by a considerable decrease in the probability of constraint satisfaction, such that $F_{LB}\approx 0.8$. 

From the analysis above, it is clear that accurate estimation of the true plant uncertainties is important if RL is to be applied to a real plant. However, it does appear that RL is likely to be robust to small amounts of error, which may be unavoidable given the limitations of finite plant data.

\section{Conclusions}\label{sec:conclusion}
Overall, in this work, we have presented an efficient RL-based methodology to address production scheduling problems, including typical constraints imposed derived from standard operating procedures and propositional logic. The methodology is demonstrated on a classical single stage, parallel batch scheduling production problem, which is modified to enable comparison between discrete-time and continuous-time formulations as well as to consider the effects of uncertainty. The RL methodology is benchmarked against a continuous-time MILP formulation. The results demonstrate that the RL methodology is able to handle constraints and plant uncertainties effectively. Specifically, on a small problem instance RL shows particular promise with performance improvements over the MILP approach by up to 4.7\%. In the larger problem instances, the RL approach had practically the same performance as MILP. Further, the RL methodology is able to optimize for various risk-aware formulations and we demonstrate the approach is able to improve the 20\% worst case performances by 2.3\% on average. The RL is orders of magnitude (150-500 times) cheaper to evaluate online than the MILP approach for the problem sizes considered. This will become even greater when larger problem sizes are considered or if compared to discrete-time scheduling formulations. Finally, the RL approach demonstrated reasonable robustness to misspecification of plant uncertainties, showing promise for application to real plants. In future work, we will consider the translation of this methodology to larger problem instances, multi-agent problems, and integration with other PSE decision-making functions, such as maintenance and process control.

\section*{Acknowledgements}
Tha authors would like to thank Panagiotis Petsagkourakis, Ilya Orson Sandoval and Zhengang Zhong for insightful discussions and feedback throughout the project. 

\bibliography{bibliography}
\appendix
\section{Particle swarm and simulated annealing (PSO-SA) hybrid algorithm}\label{app:PSO-SA}

The general intuition behind stochastic search is provided by Algorithm \ref{alg:generalSSPO}. Most stochastic search algorithms adhere to the description provided, which is summarised as follows. In \textbf{step 1} a population of parameters for a policy (network) is instantiated by a space filling procedure of choice. Then, in \textbf{step 2}, the following steps are performed \textit{iteratively}: a) a population of neural policies are constructed from the population of parameters; b\textit{i-iv}) the population is evaluated via Monte Carlo according to the estimator defined; b\textit{v}) the current best policy is tracked and stored in memory (if a criterion is satisfied); and, c) the population parameters are then perturbed according to a stochastic search optimization algorithm. After a number of optimization iterations, the best policy identified is output. The number of optimization iterations is defined according to the available computational budget or according to a threshold on the rate of improvement.

In this work, we use a hybrid particle swarm optimization - simulated annealing stochastic search optimization algorithm with a search space reduction strategy \cite{park2005particle} in order to identify the optimal scheduling policy parameters, $\theta^*\in \mathbb{R}^{n_\theta}$. Hybridisation of particle swarm optimization and simulated annealing enables us to balance exploitation and exploration of the parameter space. The search space reduction strategy was found to improve performance, especially in the \textit{harder} experiments conducted i.e. for experiment E3-8. 

\subsection{Particle Swarm Optimization}
Particle swarm optimization is a stochastic search algorithm initially proposed in \cite{kennedy1995particle}, which takes inspiration from the behaviour of animals such as fish and birds in a shoal or murmuration, respectively. The population is defined by a number of individuals, $P$. Each individual (candidate will be as a synonymous term) within the population is first initialised with a velocity, $\mathbf{v}_{i}^0\in \mathbb{R}^{n_\theta}$, and a position, $\theta_{i}^0$, $\forall i \in \{1,\ldots, P\}$. All individual's positions are subject to upper and lower bounds, such that $\theta_i = [\theta_{LB}, \theta_{UB}]^{n_\theta}$. Similarly, all velocities are subject to upper bounds, such that $\mathbf{v}_i=[-\infty, v_{max}]^{n_\theta}$, where $v_max\in \mathbb{R}$ is typically defined as:
\begin{align*}
    v_{max} = c_3(\theta_{UB}-\theta_{LB})
\end{align*}
where $c_3=[0,1]$. Each candidate is then rated with respect to an objective function, $F_{SA}: \mathbb{R}^{n_\theta} \rightarrow \mathbb{R}$ that one would like to minimize. A note of: the candidate's current best position, $\mathbf{b}^*_i\in \mathbb{R}^{n_\theta}$;  the locally best position, $\mathbf{g}_i^*\in \mathbb{R}^{n_\theta}$, within a neighbourhood of $n_h$ individuals; and, of the global best, $\theta^*$, is then made. The optimization procedure may then proceed by updating the position of each individual in the population iteratively via the following equation:
\begin{equation}
    \begin{aligned}
    \mathbf{v}_{i}^{k+1} &= \omega \mathbf{v}_i^k + c_1 r_1 \big(\mathbf{b}^*_i-\theta_i^k) + c_2 r_2(\mathbf{g}_i^* - \theta_i^k)\\
    \theta_i^{k+1} &= \theta_i^k + \mathbf{v}_i^{k+1}
    \end{aligned}\label{eq:PSOupdate}
\end{equation}
where $\omega\in \mathbb{R}_+$, $c_1\in \mathbb{R}_+$ and $c_2\in \mathbb{R}_+$ may be interpreted to represent a weighting for inertia, individual acceleration and global acceleration, respectively. The individual and local best components of the total acceleration are adjusted stochastically by the definition of $r_1\sim U(0,1)$ and $r_2\sim U(0,1)$. From Eq. \ref{eq:PSOupdate}, the update of the individual's position makes use of information both from the trajectory of the individual but the trajectories collected across the swarm.  
\subsection{Simulated Annealing}
Simulated annealing is also a stochastic search optimization algorithm, which is superficially related to structural transitions of physical systems under temperature change. The algorithm updates the position of an individual, $\theta_i^k$, probabilistically based on the improvement provided in the proposed update with respect to the objective, $F_{SA}$, and the current \textit{temperature}, $T\in \mathbb{R}_+$, of the algorithm. Specifically, an updated candidate, $\bar{\theta}_i^k$, is proposed and accepted or rejected as follows:

\begin{subequations}\label{eq:SAsystem}
    \begin{equation}\label{eq:SAperturb}
    \bar{\theta}_i^k = \theta_i^k + \mathbf{w}_i^k\\
    \end{equation}
where $\mathbf{w}_i^k=[-1,1]^{n_\theta}$ is described according to a distribution of choice on the space $[-1,1]^{n_\theta}$. Eq. \ref{eq:SAperturb} details the perturbation of a candidate's position. The evaluation and acceptance or rejection of the proposed position follows:
    \begin{equation}\label{eq:SAeval}
    \Delta E = F_{SA}(\theta_i^k) - F_{SA}(\bar{\theta}_i^k)
    \end{equation}
    \begin{equation}\label{eq:SAupdate}
    \theta_i^{k+1} =\begin{cases}
                \bar{\theta}_i^k, & \text{if } \Delta E > 0 \text{ or } \bar{z}\geq \exp(\frac{\Delta E}{T})\\
                \theta_i^k, & \text{otherwise} 
                \end{cases}
    \end{equation}
where $\bar{z}\sim U(0,1)$. From Eq. \ref{eq:SAupdate}, it is seen that if the proposed candidate does not improve with respect to the objective function, then it is accepted probabilistically to balance exploration and exploitation. 
\end{subequations}

Generally, schemes are implemented to update the temperature, $T$, given that it controls the acceptance rate. The lower the value of $T$, the higher the probability of acceptance. The larger the value of $T$ the lower the probability of acceptance. Due to the nature of the hybridisation used in this work, large constant values of $T$ were used.

\subsection{Search Space Reduction}
Search space reduction strategies are known to improve the performance of general stochastic search algorithms. In this work, preliminary experiments demonstrated that the use of a space reduction strategy improved performance of $\theta^*$ as identified by the algorithm. The strategy follows the reasoning of the work provided in \cite{park2005particle}:
\begin{equation}\label{eq:ssr}
\begin{aligned}
    \theta^{k+1}_{LB} = \theta^{k}_{LB} + \alpha (\theta^{k}_{LB} - \theta^*) \\
    \theta^{k+1}_{UB} = \theta^{k}_{UB} - \alpha (\theta^{k}_{UB} - \theta^*)
\end{aligned}
\end{equation}
where $\alpha=[0,1]$ represents a learning rate. Algorithm \ref{alg:PSO-SA} details the hybridization of the constituent components discussed in this section. 

\begin{algorithm}[h!]
\SetAlgoLined
\caption{A hybrid particle swarm optimization - simulated annealing algorithm with a search space reduction strategy}
\vspace{0.1cm}
\justify
\textbf{Input}: Initial upper, $\theta_{UB}^0$, and lower bounds, $\theta_{LB}^0$, on the parameter search space; Population size, $P$; Maximum velocity, $\mathbf{v}_{max}=[v_{max,1}, \ldots, v_{max,n_\theta}]$; Search space reduction step size, $\alpha$; Particle swarm optimization algorithm, $g_{PSO}(\cdot)$; Simulated annealing algorithm, $g_{SA}(\cdot)$; Temperature, $T$; Cooling schedule, $g_T(\cdot)$; Search space reduction rule, $g_{SSR}$; Objective function, $F_{SA}(\cdot)$; Memory buffer, $\mathcal{B}_{info}$; Maximum number of search iterations, $K$; Logic condition to trigger simulated annealing search optimization algorithm, $\text{logic}_{SA}$; Logic condition to trigger search space reduction, $\text{logic}_{SSR}$\;\justify
\textbf{1.} Generate initial population of individual parameters, $\Theta^1 = \{\theta^1_1, \ldots, \theta^1_P\}$. Each parameter setting $\theta^1_i= [\theta^1_{i,1}, \ldots, \theta^1_{i,n_\theta}] \in \Theta^1$ is generated such that, $\theta_{i,j} \sim U(\theta_{LB,j},\theta_{UB,j})$, $\forall j \in \{1, \ldots, n_\theta\}$.\vspace{0.05cm};\\
\textbf{2.} Initialize a velocity for each individual in the population via the following strategy: $\mathbf{v}^0_i = (2\mathbf{a}-1)(\theta_{UB}-\theta_{LB})$, where $\mathbf{a}=[a_1, \ldots, a_{n_\theta}]$, and $a_i \sim U(0,1)$. Define $\mathbb{V}^1 = \{\mathbf{v}_i, \forall i \in \{1, \ldots, P\}\}$\vspace{0.05cm};\\
\textbf{3.} \For{k = 1, \ldots, K}{\justify
    \textbf{a.} Evaluate $J_i^k=F_{SA}(\theta_i),\text{ }\forall \theta_i \in \Theta^k$. Append $J_i$ to $\mathcal{B}_{info}, \text{ } \forall i$\;\justify
    \textbf{b.} Determine the current local best, $\mathbf{g}^*_i$, for a neighbourhood composed of $n_h$ individuals, for $\theta_i \in \Theta^k$. Define $\mathbb{G}^k=\{\mathbf{g}^*_i, \forall i \in \{1, \ldots, P\}\}$.\;\justify
    \textbf{c.} According to $J_i^{j}\in\mathcal{B}_{info}, \text{ } \forall j \in\{1, \ldots, k\}$, determine whether $\theta^k_i$ is the best parameter setting observed by the individual. Store this position as the current individual best, $\mathbf{b}_i^*$ and store in $\mathbb{B}^k =\{\mathbf{b}^*_i, \forall i \in \{1, \ldots, P\}\}$\;\justify
    \textbf{d.} \If{$\text{logic}_{SSR}$}{
                \textbf{i.} Reduce search space: $\theta^k_{UB}, \theta^k_{LB} = g_{SSR}(\mathbb{G}^k, \mathbb{\theta}^{k})$}\justify
    \textbf{e.} \If{$\text{logic}_{SA}$}{\justify
                    \textbf{i.} Conduct simulated annealing and update individual and neighbourhood best: $\Theta^{k}, \mathbb{G}^k, \mathbb{B}^k, \mathcal{B}_{info} = g_{SA}(\Theta^{k}, \mathbb{G}^k, \mathbb{B}^k, \mathcal{B}_{info}, F_{SA}, T^k)$, via e.g. Eq. \ref{eq:SAsystem}\;\justify
                    \textbf{ii.} Update temperature: $T^{k+1} = g_T(T^k)$ }\justify
    \textbf{f.} Update population via Particle Swarm: $\Theta^{k+1}, \mathbb{V}^{k+1} = g_{PSO}(\Theta^{k}, \mathbb{G}^k, \mathbb{B}^k)$ via e.g. Eq. \ref{eq:PSOupdate}. Ensure $\theta_i^{k+1}\in\Theta^{k+1}$ and $\mathbf{v}^{k+1}_i \in \mathbb{V}^{k+1}$ satisfy constraints provided by $\theta_{LB}$, $\theta_{UB}$ and $\mathbf{v}_{max}$\;}
\justify\textbf{4.} Determine global best parameters, $\theta^*$, from $\mathbb{B}^K$ and $\mathcal{B}_{info}$\;%
\justify\textbf{Output:} $\theta^*$ \vspace{0.1cm}
\label{alg:PSO-SA}
\end{algorithm} 
\FloatBarrier

\subsection{Policy network structure selection}

Generally, it was observed that the performance of $\pi$ learned via the method proposed was dependent upon proper selection of the neural network structure. It was found that a neural network 3 layer network with: an input layer of $n_x= 2N + 2u +1$ nodes, where $N$ is the number of tasks and $n_u$ is the number of units; hidden layer 1 composed of 10 feedforward nodes; hidden layer 2 composed of 4 Elman recurrent (tanh) nodes; hidden layer 3 composed of feedforward 2 nodes; and, an output layer composed of $n_u$ nodes. Hyperbolic tangent activation functions were applied across hidden layer 2, a sigmoid over hidden layer 3, and a ReLU6 function over the output layer. The investigation of deeper networks was generally led by research supporting their use for approximation of non-smooth functions, as provided in \cite{imaizumi2018deep}. The use of recurrency within the network was used to handle the potnetial partial observability of the problem when uncertainty was present in case study \cite{heess2015memory}. 

\section{Definition of the production scheduling problem}\label{app:model}
The following text defines the problem case studies presented in this work. Full code implementation of the system defined here is available at \url{https://github.com/mawbray/ps-gym} and is compatible with stable RL baseline algorithms (as it is coded via the recipe provided by OpenAI gym custom classes).

\subsection{Problem definition}
We consider a multi-product plant where the conversion of raw material to product only requires one processing stage. \textit{We assume} there is an unlimited amount of raw material, resources, storage and wait time (of units) available to the scheduling element. Further, we assume that the plant is completely reactive to the scheduling decisions of the policy, $\pi$, although this assumption can be relaxed if decision-making is formulated within an appropriate framework as shown in \cite{hubbs2020deep}. The scheduling element \textit{must decide} the sequencing of tasks (which correspond uniquely to client orders) on the equipment (units) available to the plant. The following operational rules are imposed on the scheduling element: 
\begin{enumerate}\label{list}
    \item A given unit \textit{l} has a maximum batch size for a given task \textit{i}. Each task must be organized in campaigns (i.e. processed via multiple batches sequentially) and completed once  during the scheduling horizon. All batch sizes are predetermined, but there is uncertainty as to the processing time (this is specific to task and unit).
    \item Further, the task should be processed before the delivery due date of the client, which is assumed to be an uncertain variable (the due date is approximately known at the start of the scheduling horizon, but is confirmed with the plant a number of periods before the order is required by the client).
    \item There are constraints on the viable sequencing of tasks within units (i.e. some tasks may not be processed before or after others in certain units).
    \item There is a sequence and unit dependent cleaning period required between operations, during which no operations should be scheduled in the given unit.
    \item Each task may be scheduled in a subset of the available units.
    \item Some units are not available from the beginning of the horizon and some tasks may not be processed for a fixed period from the start of the horizon (i.e. they have a fixed release time).
    \item Processing of a task in a unit must terminate before another task can be scheduled in that unit.
\end{enumerate}

The objective of operation is to minimize the makespan and the tardiness of task (order) completion. This means that once all the tasks have been successfully processed according to the operational rules defined, then the decision making problem can be terminated. As in the original work, we formalize the notation of a number of problem sets and parameters in Table \ref{table:sets_params}. We try to keep notation consistent with that work. It should be noted that in this work, we transcribe the problem as a discrete-time formulation. The original work \cite{cerda1997mixed} utilized a continuous-time formulation. Further discussion is provided later in the text on this topic. %
We formalize the notation of a number of problem sets and parameters in Table \ref{table:sets_params}.

\begin{table}[h!]
  \caption{Table of problem parameters and sets. *D.T.I. is shorthand for discrete time indices.}
  \label{table:sets_params}
  \small
  \centering
  \begin{tabular}{ll}
    \toprule
    Sets & Notation \\
    \midrule
    Tasks (orders) to be processed & $\mathbb{I} = \{1, \ldots, N\}\subset \mathbb{Z}_+$ \\
    Available units & $\mathbb{L} = \{1, \ldots, n_u\}\subset \mathbb{Z}_+$\\
    Available units for task \textit{i} & $\mathbb{L}_i\subseteq \mathbb{L}$ \\
    The task most recently processed in unit \textit{l}& $\mathbb{M}_l \subset \mathbb{Z}$; $|\mathbb{M}_l|\leq1$\\
    Tasks which have been completely processed & $\mathbb{T}_f\subset \mathbb{Z}$ \\
    Feasible successors of task \textit{i} in unit \textit{l} & $\mathbb{SU}_{il}\subset \mathbb{Z}_+$ \\
    Feasible successors of task \textit{i} & $\mathbb{SU}_{i} = \cup_{l\in \mathbb{L}} \mathbb{SU}_{il} \subset \mathbb{Z}_+ $\\
    The task currently being processed in unit \textit{l} & $\mathbb{O}_l$\\
    \toprule
    Parameters & Notation  \\
    \midrule
    Number of tasks & $N\in \mathbb{R}_+$\\
    Due date of client order (task) & $\tau_i\in \mathbb{Z}_+$  \\
    Number of units & $n_u \in \mathbb{Z}_+$  \\
    Batch size of unit \textit{l} for task \textit{i} & $B_{il}\in \mathbb{R}_+$ \\
    Number of batches required to process task \textit{i} in unit \textit{l} & $NB_{il}\in \mathbb{Z}_+$ \\ 
    Sequence dependent set up time  & $\hat{\upsilon}_{ilt}\in \mathbb{Z}_+$ \\
    Release time of tasks in D.T.I. & $(\text{RTT})_i\in \mathbb{Z}_+$  \\
    Release time of units in D.T.I. & $(\text{RTU})_l\in \mathbb{Z}_+$ \\
    Cleaning time required for unit \textit{l} between processing tasks \textit{m} and \textit{i} successively & $TCL_{mil}\in \mathbb{Z}_+$ \\
    \toprule
    Miscellaneous Variables & Notation \\
    \midrule
    Variable indicating campaign production of task \textit{i} starts at time \textit{t} in unit \textit{l} & $W_{ilt} \in \mathbb{Z}_2$ \\
    Variable indicating unit \textit{l} is processing task \textit{i} at time \textit{t} & $\bar{W}_{ilt} \in \mathbb{Z}_2$ \\
    Integer variable denoting task \textit{i} is being processed in unit \textit{l} at time \textit{t} & $w_{lt} \in \mathbb{U}$ \\
    Estimated D.T.I. for unit \textit{l} to process a batch of task \textit{i} & $\bar{PT}_{il} \in \mathbb{Z}$ \\
    Actual D.T.I. for unit \textit{l} to process batch $n$ of task \textit{i} & $PL_{inl} \in \mathbb{Z}$ \\
    Actual D.T.I. to finish processing current campaign in unit \textit{l} at time \textit{t}   & $\delta_{lt} \in \mathbb{Z}$ \\
    Current inventory of task (client order) \textit{i} at time \textit{t} & $I_{it}\in \mathbb{R}_+$ \\
    D.T.I. until due date of task \textit{i} at time \textit{t} & $\rho_{it}\in \mathbb{Z}$ \\
    The length of a discrete time index & $dt\in \mathbb{R}_+$ \\
    \bottomrule
  \end{tabular}
\end{table}
\subsection{Formulating discrete-time scheduling problems as Markov decision processes}
The methodology presented in Section \ref{sec:methodology} utilizes formulation of the scheduling problem as an MDP, which is an approximation to solving the finite-horizon stochastic optimal control problem (as detailed in Eq. \ref{eq:CSOCP}). Construction of the problem follows.

Firstly, the time horizon is discretized into $T=200$ finite periods of length $dt=0.5$ days, where $t \in \{0, \ldots, T\}$ describes the time at a given index in the discrete time horizon. 

We hypothesise that the system is made completely observable by the following state representation:
\begin{equation}\label{eq:x_t}
    \mathbf{x}_{t} = \bigg[I_{1 t}, \ldots, I_{N t}, w_{1 t}, \ldots, w_{n_u t},\delta_{1 t}, \ldots, \delta_{n_u t}, \rho_{1 t}, \ldots, \rho_{N t}, t\bigg] ^T\in \mathbb{R}^{2N +2n_u + 1}
\end{equation}
where $I_{i t} \text{ }\forall i\in \mathbb{I}$ quantifies the current inventory of client orders in the plant; the tasks processed within units in the plant over the previous time interval, $w_{l t} \text{ } \forall l\in \mathbb{L}$; the discrete time indices remaining until completion of the task campaigns being processed in all units, $\delta_{lt} \text{ } \forall l \in \mathbb{L}$; the discrete time indices remaining until orders (tasks) are due, $\rho_{it}\text{ }\forall i\in \mathbb{I}$; and, the current discrete time index, $t$.

Similarly, we define the control space, $\mathbf{u} = [u_1, \ldots, u_{n_u}]^T \in \mathbb{U}$ as the set of integer decisions (tasks) that could be scheduled in the available units. Hence, one can define $\mathbb{U} = \bigcup_{l=1}^{n_u} \mathbb{U}_{l} \subset \mathbb{Z}^{n_u}_+$, where $\mathbb{U}_{l} = \mathbb{I}\cup \{N+1\}$, and $N+1$ is an integer value, which represents the decision to idle the unit for one time period. 

A sparse reward function is defined as follows:
\begin{equation}\label{eq:RF}
    R =
    \begin{cases}
     \mathbf{d}^T\mathbf{x}_{t+1}, & \text{if } t = T-1 \text{ or } \mathbb{T}_f = \mathbb{I} \\
     0, & \text{otherwise}
    \end{cases}
\end{equation}
where $\mathbf{d} \in \mathbb{Z}^{n_x}$ denotes a vector specified to penalise order lateness and makespan; $\mathbb{I}\subseteq \mathbb{Z}$ denotes the set of tasks (or client orders); and, $\mathbb{T}_f\subseteq \mathbb{Z}$ denotes the set of tasks, which have been completed. If $\mathbb{T}_f = \mathbb{I}$ is satisfied, the decision-making problem is terminated. Definition of $\mathbf{d}$ follows:
\begin{equation}\label{eq:rfc}
    \mathbf{d} = \big[\mathbf{0}_{1,N}, \mathbf{0}_{1,n_u}, \mathbf{0}_{1,n_u}, \mathbf{1}_{1,N}, -1\big]^T\in \mathbb{Z}^{2N+2n_u+1}
\end{equation}

where $\mathbf{0}_{1,j}$ represents a row vector of zeros of dimension $j$; and, $\mathbf{1}_{1,j}$ represents a row vector of ones of dimension $j$. How the definition provided by Eq. \ref{eq:RF} enforces operational objectives is clarified by full definition of the dynamics, as presented subsequently. 

The inventory balance is defined to highlight that the underlying state space model is both non-smooth and subject to uncertain variables:
\begin{equation}\label{eq:inventory_balance}
    \begin{aligned}
    I_{it+1} &= I_{it} + \sum_{n=1}^{NB_{il}}\sum_{l\in \mathbb{L}_i} B_{il} W_{ilt-\hat{\upsilon}_{ilt}-\sum_{\hat{n}=1}^{n}PL_{i\hat{n}l}}  \quad \forall i \in \mathbb{I}
    \end{aligned}
\end{equation}
where ${B}_{il}$ is the maximum batch size (kg) of unit $l$ for task \textit{i}; ${NB}_{il}$ is the integer number of batches required to satisfy the client's order size in unit \textit{l}; ${PL}_{inl}$ is a random variable that indicates the number of discrete time indices required to produce batch \textit{n} of task \textit{i} in unit \textit{l} and represents a realization of an uncertain variable; $W_{ilt}\in \mathbb{Z}_2$ is a binary variable, indicating a production campaign for task \textit{i} was first scheduled in unit \textit{l} at time \textit{t}; $\hat{\upsilon}_{ilt}$ defines the current sequence dependent unit set up time required to process order \textit{i} in unit \textit{l} at time \textit{t}. The sequence dependent unit set up time is defined as follows:
\begin{align}
    \hat{\upsilon}_{ilt} &= \begin{cases} \sum_{m\in \mathbb{M}_l}\max(\max([\text{TCL}_{mil}+t_{ml}-t]^+, [(\text{RTT})_i -t]^+),[(\text{RTU})_l-t]^+\big), & \text{if } |\mathbb{M}_l| = 1 \vspace{0.2cm} \\
       \max([(\text{RTT})_i -t]^+,[(\text{RTU})_l-t]^+\big)   ,& \text{otherwise}
        \end{cases} \\
\end{align}
where $\mathbb{M}_l \subset \mathbb{Z}_+$, defines a set denoting the task most recently processed in unit \textit{l}, which finished at time, $\textit{t}_{ml} \leq t$, such that the cardinality of the set $|\mathbb{M}_l| \in \mathbb{Z}_2$; $\text{TCL}_{mil}\in \mathbb{R}_+$ defines the cleaning times between successive tasks \textit{m} and \textit{i}; $(\text{RTT})_i\in \mathbb{Z}_+$ defines the release time (in number of discrete time indices) a task \textit{i} cannot be processed for at the start of the horizon; $(\text{RTU})_l\in \mathbb{Z}_+$ defines the release time of unit \textit{l}.

It is apparent that in order to handle the requirements for cleaning time between processing of successive tasks \textit{m} and \textit{i} in unit \textit{l}, we directly incorporate the cleaning time required and all other mandated setup time (collectively denoted $\hat{\upsilon}_{ilt}$) into the total processing time of task \textit{i} in unit \textit{l} when first scheduled at time \textit{t}. The total processing time for a task \textit{i} in unit \textit{l} at time $\hat{\textit{t}}_l$ is then equivalent to $\sum_{n=1}^{NB_{il}} PL_{inl} + \hat{\upsilon}_{il\hat{\textit{t}}_l}$. 

Further, if a different task $\hat{\textit{i}}$ is scheduled by the policy at time, $\bar{\textit{t}}$, in unit \textit{l}, before the end of the production campaign for task \textit{i} (i.e. where $W_{il\hat{t}_l}(\sum_{n=1}^{NB_{il}}PL_{inl}+\hat{\upsilon}_{il\hat{t}_l})+\hat{t}_l>\bar{\textit{t}}> \hat{t}_l$),  then the task indicator $W_{il\hat{t}_l}= 0$ is redefined. 

Similarly, if a production campaign of task \textit{i} successfully terminates at time \textit{t} in unit \textit{l}, then:  $W_{il\hat{t}_l}= 0$; $\mathbb{M}_l = \{i\}$; $t_{ml} = t$; and, the task is added to a list of completed orders, $\mathbb{T}_f = \mathbb{T}_f\cup \{\textit{i}\}$. In both cases redefinition of the binary variable is retroactive to the initial scheduling decision.

Next, we update a representation of 'lifting variables' \cite{gupta2017general} that indicate current unit operations:
\begin{equation}\label{eq:appunit_processing}
    w_{lt+1} = u_{lt} \quad \forall l \in \mathbb{L}
\end{equation}

where $u_{lt} \in \mathbb{Z}_+$ is the scheduled decision at the previous time index. This is deemed a necessary part of the state representation, given that it provides information regarding the operational status of the available equipment at the current time index. For example: if $u_{lt} = N+1$, then the unit is idle; if $0 \leq u_{lt} < N+1$, then it is not. A control must be predicted at each discrete time index. Typically, the length of a production campaign will be greater than the duration of discrete time interval, $dt$. To handle this if ${u}_{lt}={u}_{lt-1}$ it is deemed that the scheduling element is indicating to continue with the current operation in unit \textit{l}. 

To determine the amount of processing time left for a given campaign in a unit \textit{l}, we define:
\begin{equation}\label{eq:appunit_task}
\begin{aligned}
    A_{lt} &= \sum_{i\in \mathbb{I}_L}W_{ilt}\bigg(\sum_{n=1}^{NB_{il}} PL_{inl} + \hat{\upsilon}_{ilt} -\delta_{lt}\bigg) + \delta_{lt}\\
    \delta_{lt+1} &= A_{lt} - 1  \quad \forall l \in \mathbb{L} \\
\end{aligned}
\end{equation}
The formulation assumes that we know the processing time required for batch $n$ of task $i$ in unit $l$, $PL_{i,n,l}$, ahead of its realization. In reality it is treated as a random variable, which is only observed when a given batch in a production campaign terminates. This is discussed further in Section \ref{sec:uncertainobs}. The final equation comprising the system updates the number of discrete time indices until the order \textit{i} is to be completed for delivery to the client, $\rho_{it}$:
\begin{equation}\label{eq:appdue_date}
    \rho_{it+1} = \begin{cases}
    [\rho_{it}]^-, & \text{if } i \in \mathbb{T}_f\\
    \rho_{it} - 1, & \text{otherwise}
    \end{cases} \quad \forall i \in \mathbb{I}
\end{equation}
where $[y]^- = \min(0,y)$. The use of the logic condition enforced in Eq. \ref{eq:appdue_date} is essentially redundant from the point of view of providing information about completed tasks, given that this is equivalently expressed by the inventory representation in the state vector. However, it is particularly useful in the allocation of rewards as it provides means to easily allocate penalty for tardy completion of tasks as expressed in Eq. \ref{eq:RF}.

\subsection{A forecasting framework for handling future plant uncertainty}\label{sec:uncertainobs}
In this study, the processing time of a batch, $PL_{inl}$, and the due date of the client orders, $\tau_i$, are described by uniform and poisson distributions, respectively. This is detailed bu Eq. \ref{eq:uniformpt} and \ref{eq:poissondue_date}:
\begin{subequations}
\begin{equation}\label{eq:uniformpt}
\begin{aligned}
    PL_{inl} &\sim U(\max(1,\bar{PT}_{il}-c), \bar{PT}_{il}+c)  \quad  \forall n \in \{1, \ldots,NB_{il}\}, \quad \forall l \in \mathbb{L}_i, \quad \forall i \in \mathbb{I}\\
\end{aligned}
\end{equation}
where $c\in \mathbb{Z}_+$ defines the variance of the distribution ($c=1$ in this work); and, $\bar{PT}_{il}$ is the expected processing time for a batch of task \textit{i} in unit \textit{l}. The due date uncertainty is described as follows: 
\begin{equation}\label{eq:poissondue_date}
    \tau_i \sim P(\bar{\tau}_i) \quad \forall i \in \mathbb{I}
\end{equation}
where $\bar{\tau}_i$ is the expected due date. In practice, we do not observe the realizations of these variables in advance, and hence maintain estimates according to their expected values within the state representation (i.e. Eq. \ref{eq:x_t}).  For example, in the case of the due date, the client is required to confirm the delivery date a number of days before the order - at which point $\tau_i$ is observed and updated within the state.
\end{subequations}

\subsection{Defining the initial system state}
Eq. \ref{eq:inventory_balance} - \ref{eq:appdue_date} represent the discrete-time plant dynamics (i.e. Eq. \ref{eq:plantdyn}) that we are concerned with. To initiate the system, it is necessary to define an initial state, $\mathbf{x}_0$, that represents the plant at the start of the scheduling horizon, $t=0$. This is described by the following set of expressions:
\begin{subequations}
\begin{equation}\label{eq:appinitinventory_balance}
    I_{i0} = 0 \quad \forall i \in \mathbb{I}
\end{equation}
which assumes that at the start of the horizon, the plant holds no inventory of the products one desires to produce;
\begin{equation}\label{eq:appinitunit_processing}
    w_{l0} = N+1 \quad \forall l \in \mathbb{L}
\end{equation}
denotes that at the start of the horizon all units are idled. Even if this is not the case (i.e. the unit is unavailable for other reasons, such as production of another task which it began before the start of the scheduling horizon), the allocation appears satisfactory according to experiments that follow in Section \ref{sec:R&D};
\begin{equation}\label{eq:appinitunit_task}
\begin{aligned}
    \delta_{l0} &= (\text{RTU})_l  \quad \forall l \in \mathbb{L} \\
\end{aligned}
\end{equation}
Eq. \ref{eq:appinitunit_task} assumes that the unit is required to be idled for at least a period equivalent to its release time. At the start of the horizon, the number of discrete time indices until a task is due to be delivered to a client is equivalent to the task due date:
\begin{equation}\label{eq:appinitdue_date}
    \rho_{i0} = \bar{\tau}_i \quad \forall i \in \mathbb{I}
\end{equation}\label{eq:initsystem}
\end{subequations}
It should also be noted that at $t=0$, $\mathbb{M}_l = \{\}\text{ } \forall l$ and $\mathbb{T}_f=\{\}$. 

Having provided detail of the state, controls space and plant dynamics, we now turn our attention to identification of the feasible controls. 

\subsection{Defining the set of feasible controls}\label{app:constraintset}
We seek to identify the set of controls $\hat{\mathbb{U}}_t \subset \mathbb{U} \subset \mathbb{Z}^{n_u}$, which innately satisfy the constraints imposed on the scheduling element. 

Here, we define a number of sets that enable us to identify $\bar{\mathbb{U}}_t$:
\begin{enumerate}
    \item A given unit \textit{l} can process a subset, $\mathbb{I}_l \subseteq \mathbb{I}$ of the tasks that require processing.
    \item If unit \textit{l} has just processed task \textit{i} then there exists a set of tasks which may be processed subsequently, $\mathbb{SU}_{il}$.
    \item There exists a set, $\mathbb{O}_l = \{i\}$, which describes the task currently being processed in unit \textit{l}. At $t=0$, $\mathbb{O}_l = \emptyset$. If the unit is idled, then $\mathbb{O}_l = \emptyset$. 
    \item As previously mentioned there exists a set, $\mathbb{T}_f$, descriptive of those tasks, which have already been processed successfully. At $t=0$, $\mathbb{T}_f = \emptyset$.
    \item Lastly, again, there exists a set, $\mathbb{M}_l$ descriptive of the task most recently processed in unit \textit{l}. Likewise, at $t=0$, $\mathbb{M}_l = \emptyset$.
\end{enumerate} 
It follows then that any given time in processing, one can define the set of controls that partially satisfy the imposed operational rules as: 
\begin{equation}\label{eq:appautouhat}
    \begin{aligned}
    \bar{\mathbb{U}}_t = \bigcup_{l=1}^{n_u} \begin{cases}
     (\mathbb{I}_{l}\cup \{N+1\}) \backslash \mathbb{T}_f & \text{if } \mathbb{O}_l = \emptyset \text{ \& } \mathbb{M}_l = \emptyset \\
    ((\mathbb{I}_{l}\cup \{N+1\})\cap\mathbb{SU}_{ml})\backslash \mathbb{T}_f \text{ where, } m\in \mathbb{M}_l & \text{if } \mathbb{O}_l = \emptyset \text{ \& } |\mathbb{M}_l| = 1 \\
    \mathbb{O}_l & \text{ if } \mathbb{O}_l \neq \emptyset
    \end{cases}
    \end{aligned}
\end{equation}
Note, however, in this instance we cannot innately satisfy the control set $\hat{\mathbb{U}}_t$, as $\hat{\mathbb{U}}_t \subset \bar{\mathbb{U}}_t$. Specifically, $\bar{\mathbb{U}}_t$ permits the scheduling of the same task \textit{i} in two different units \textit{l} and \textit{l'} at the same time index \textit{t}. We propose to handle this through use of a penalty function, which we define as:
\begin{equation}\label{eq:appPFCS}
    \begin{aligned}
    \phi &= R - \kappa_g\left\lVert [g(\mathbf{x},\mathbf{u})]^+ \right\rVert_2\\
    g(\mathbf{x}, \mathbf{u}) &= [g_1, \ldots, g_N]\\
    g_i &= \sum_{l \in \mathbb{L}_i} \bar{W}_{ilt}-1\\
    \end{aligned}
\end{equation}
where $\kappa_g = 250$ and $\bar{W}_{ilt}\in \mathbb{Z}_2$ is a lifting variable that indicates unit \textit{l} is processing task \textit{i} at time \textit{t}, such that if $u_{lt} = i$, then $\bar{W}_{ilt}=1$ and, otherwise $\bar{W}_{ilt}=0$. Note that technically, under the discrete-time state space model formulation used in this work (i.e. Eq. \ref{eq:appunit_processing}), this could also be considered as a state inequality constraint. However, due to the uncertainty associated with the evolution of the state, this formalism will be explored in future work.
By deploying the methodology proposed in Section \ref{sec:methodology}, but modifying the rounding policy such that it is defined $f_r: \mathbb{W} \rightarrow \bar{\mathbb{U}}$, we can ensure that the decisions of the policy function, $\pi$ satisfy the constraints imposed on the scheduling problem (see \ref{list}) as originally proposed in \cite{cerda1997mixed} by maximising the \textit{penalised return} defined under $\phi$.

\section{Definition of experimental data used in computational experiments}\label{app:data}

This section details the data used to define the case studies investigated in Section \ref{sec:R&D}. Please see Tables \ref{table:maxbatch}, \ref{table:processingtimesbatch}, \ref{table:sequenceviable}, \ref{table:cleaningtimes}, \ref{table:ordersizeduedatert} for information regarding: the feasible processing of tasks in units and respective maximum batch sizes; nominal processing times; the viable successors of a given task; the cleaning times required between successive tasks; and information regarding order sizes, due dates and release times, respectively. 

\begin{table}[h!]
  \caption{Maximum task batch size (kg/batch) for every unit. RTU\textsuperscript{*} denotes the finite release time of the unit in days. The length of a discrete time index corresponds to 0.5 days.}
  \label{table:maxbatch}
  \small
  \centering
  \begin{tabular}{ccccc}
    \toprule
    & \multicolumn{4}{c}{Unit, \textit{l}} \\
    \midrule
    Task (order) & 1 & 2 & 3 & 4\\
    \midrule
    T1 & 100&       &   &  \\
    T2 &    &       & 210&\\
    T3 & 140&       & 170&\\
    T4 &    & 120   & &\\
    T5 &    & 90    & & 130 \\
    T6 & 280 & 210  & & \\
    T7 &    &       & 390&290\\
    T8 &    &       & & 120 \\
    T9 & 200 &      & &\\
    T10 & 250 &  270& & \\
    T11 &   &       & 190& \\
    T12 &   &       & 140& 150\\
    T13 & 120 &     & 155 &\\
    T14&    &   115 & & \\
    T15 &   &   130 & & 145 \\
    RTU\textsuperscript{*} & 0.0 & 3.0 & 2.0 & 3.0 \\
    \bottomrule
  \end{tabular}
\end{table}

\begin{table}[h!]
  \caption{Order processing times (days/batch), $PT_{il}$. The length of a discrete time index corresponds to 0.5 days.}
  \label{table:processingtimesbatch}
  \small
  \centering
  \begin{tabular}{ccccc}
    \toprule
    & \multicolumn{4}{c}{Unit, \textit{l}} \\
    \midrule
    Task (order), $\text{T}_i$ & 1 & 2 & 3 & 4\\
    \midrule
    T1 & 2.0&       &   &  \\
    T2 &    &       & 1.0&\\
    T3 & 1.0&       & 1.0&\\
    T4 &    & 1.5   & &\\
    T5 &    & 1.5    & & 1.0 \\
    T6 & 2.5 & 2.0  & & \\
    T7 &    &       & 1.0& 1.5\\
    T8 &    &       & &  2.0 \\
    T9 & 1.5 &      & &\\
    T10 & 2.5 &  2.0& & \\
    T11 &   &       & 1.5& \\
    T12 &   &       & 2.0& 1.5\\
    T13 & 3.0 &     & 1.0 &\\
    T14&    &   2.5 & & \\
    T15 &   &   1.0 & & 2.0 \\
    \bottomrule
  \end{tabular}
\end{table}

\begin{table}[h!]
  \caption{Set of feasible successors.}
  \label{table:sequenceviable}
  \small
  \centering
  \begin{tabular}{cc}
    \toprule
    Task (order) & Feasible Successors\\
    \midrule
    T1 & T6, T9, T10, T13  \\
    T2 &  T3, T11, T13\\
    T3 & T1, T2, T7, T9, T10, T11, T12\\
    T4 &  T5, T10, T14, T15 \\
    T5 & T4, T6, T7, T8, T12, T14 \\
    T6 & T1, T3, T4, T9, T13, T14, T15\\
    T7 & T2, T5, T8, T12 \\
    T8 & T7, T11, T12, T15  \\
    T9 & T1, T3, T6, T10, T13\\
    T10 & T1, T3, T4, T9, T13, T14, T15\\
    T11 &  T2, T12, T13 \\
    T12 & T3, T5, T7, T8, T11, T15\\
    T13 & T1, T2, T3, T6, T7, T9, T10, T11, T12\\
    T14 &  T4, T5, T6, T10, T15   \\
    T15 &  T4, T6, T7, T8, T10, T12, T14  \\
    \bottomrule
  \end{tabular}
\end{table}

\begin{table}[h!]
  \caption{Cleaning times required between pairs of orders (days) in all units. The length of a discrete time index corresponds to 0.5 days.}
  \label{table:cleaningtimes}
  \small
  \centering
  \begin{tabular}{cccccccccccccccc}
    \toprule
    & \multicolumn{15}{c}{Succeeding Task} \\
    \midrule
    Task (order) & T1 & T2 & T3 & T4 & T5 & T6 & T7 & T8 & T9 & T10 & T11 & T12 & T13 & T14 & T15 \\
    \midrule
    T1 &  &  &  &  &  & 0.5  &  &  & 1.0 & 0.5 &  &  &1.5  &  &  \\
    T2 &  &  & 1.0 &  &  &   &  &  &  &  & 1.0 &  & 1.5 &  & \\
    T3 & 1.0 & 0.5 &  &  &  &   & 0.5 &  & 1.5 & 0.5 & 1.0 & 2.0 &  &  & \\
    T4 &  &  &  &  & 0.5 &   &  &  &  & 0.5  &  &  &  & 2.0 & 1.0 \\
    T5 &  &  &  & 0.5 &  &  0.5 & 1.0 & 0.5 &  &  &  & 0.5 &  & 0.5 & \\
    T6 & 1.5 &  & 0.5 & 0.5 &  &   &  &  &1.0  &  &  &  &  0.5 & 1.0 & 1.5\\
    T7 & & 2.0 &  &  & 1.0 &   &  & 0.5 &  &  &  & 1.0 &  &  & \\
    T8 & &  &  &  &  &   & 1.5  &  &  &  & 0.5 & 0.5 &  &  & 1.5 \\
    T9 & 2.0 &  & 1.0 &  &  &  0.5 &  &  &  & 1.5 &  &  & 3.0 &  & \\
    T10 & 1.0 &  & 0.5  & 0.5 & 1.0  &   &  &  & 2.5 &  &  &  & 0.5  & 2.0 & 1.0 \\
    T11 & & 1.0  &  &  &  &   &  &  &  &  &  & 0.5 & 2.5 &  & \\
    T12 &  &  & 1.0 &  & 1.5 &   & 2.0 & 1.0 &  &  & 0.5 &  &  &  & 1.0 \\
    T13 & 1.5 & 0.5 & 2.0 &  &  & 2.0 & 2.5 &  & 0.5 & 0.5 & 1.0 & 1.5  &  &  &  \\
    T14& &  &  & 0.5 & 0.5 & 0.5 &  &  &  & 0.5 &  &  &  &  & 0.5 \\
    T15 & &  &  & 0.5 &  & 0.5 & 0.5 & 0.5 &  & 1.5 &  & 0.5 &  & 0.5 &  \\
    \bottomrule
  \end{tabular}
\end{table}

\begin{table}[h!]
  \caption{Order sizes, due dates (days) and release times. The length of a discrete time index corresponds to 0.5 days.}
  \label{table:ordersizeduedatert}
  \small
  \centering
  \begin{tabular}{cccc}
    \toprule
    Task (order) $\text{T}_i$ & Order size, $M_i$ (kg) & Due date, $\bar{\tau}_i$ (days) & Release time of order, $(RTO)_i$ (days) \\
    \midrule
    T1 & 700 & 10 & 0 \\
    T2 & 850 & 22 & 5 \\
    T3 & 900 & 25 & 0\\
    T4 & 900 & 20 & 6\\
    T5 & 500 & 28 & 0 \\
    T6 & 1350 & 30 & 2\\
    T7 & 950 & 17 & 3\\
    T8 & 850 & 23 & 0\\
    T9 & 450 & 30 & 2\\
    T10 & 650 & 21 & 6\\
    T11 & 300 & 30 & 0\\
    T12 & 450 & 28 & 1.5\\
    T13 & 200 & 15 & 0\\
    T14& 700 & 29 &  0\\
    T15 & 300 & 40 & 5.5\\
    \bottomrule
  \end{tabular}
\end{table}
\FloatBarrier

\section{Misspecification of plant uncertainty}\label{app:uncertaintymisspec}

The processing time uncertainty has the same form as Eq. \ref{eq:uniformpt}, but with misspecification of the parameters of the distribution by an additive constant, $k_{pt}\in \mathbb{Z}_+$. In this case, the actual plant processing time uncertainty, Eq. \ref{eq:uniformpt}, is redefined:
\begin{align*}
    PL_{inl} \sim U\big(\max(1, \bar{PT}_{il} - \hat{c}), \bar{PT}_{il} + \hat{c}\big), \quad \forall n\in\{1,\ldots, NB_{i,l}\}, \quad \forall l \in \mathbb{L}_i, \quad \forall i \in \mathbb{I} 
\end{align*}
where $\hat{c} = c +k_{pt}$ (in this work $c=1$). Similarly, the rate of the Poisson distribution descriptive of the due date uncertainty (Eq. \ref{eq:poissondue_date}) is misspecified. Here, however, the misspecification is treated probabilistically, such that the rate, $\bar{\tau}_i, \text{ } \forall i\in \mathbb{I}$, is perturbed by a relatively small amount. Specifically, we redefine the due date uncertainty of the real plant as follows:
\begin{align*}
    \tau_i \sim P(\hat{\tau}_i), \quad \forall i \in \mathbb{I}
\end{align*}
where $\hat{\tau}_i = \bar{\tau}_i + k_{dd}z_i , \text{ } z_i\sim U(-1,1) \text{ } \forall i\in \mathbb{I} \text{ and } k_{dd}\in \mathbb{Z}_+$. Details of $k_{pt}$ and $k_{dd}$ investigated in this work are provided by Table \ref{table:exp_misspec} in Section \ref{sec:robustRL}.

\section{The probability of constraint satisfaction}\label{app:constraintssamples}
In this work, we are interested in satisfying hard constraints on the scheduling decisions (control inputs) to a plant. In the case that the underlying plant is subject to uncertainties, we evaluate this constraint as we would a soft constraint, i.e. probabilistically. This means we are interested in quantifying the probability, $F_U$, that the constraints on the control inputs, $\mathbf{u}_t \in \hat{\mathbb{U}}_t, \text{ } \forall t$ are satisfied:
\begin{equation}
\begin{aligned}
    F_{U} = \mathbb{P}\bigg(\bigcap_{i=0}^{T-1}\{\mathbf{u}_i \in \mathbb{\hat{U}}_i\}\bigg) 
\end{aligned}
\end{equation}
In order to estimate $F_U$, we gain empirical approximation, $F_{SA}$, known as the empirical cumulative distribution function, by sampling. In this work, $n_I=500$ Monte Carlo samples were used to estimate $F_{LB}$ in the results reported as follows:
\begin{equation}
    F_{SA} = \frac{1}{n_I}\sum_{i=1}^{n_I} Y^i
\end{equation}
where $Y^i$ is a random variable, which takes a value of $Y^i=1$, if $\mathbf{u}_t \in \hat{\mathbb{U}}_t\text{, } \forall t$, and $Y^i=0$ if $ \exists \text{ } \mathbf{u}_t \notin \hat{\mathbb{U}}_t\text{, } \forall t$. Due to the nature of estimation from finite samples, the $F_{SA}$ is prone to estimation error.Hence, we employ a concept from the \textit{binomial proportion confidence interval} literature, known as the Clopper–Pearson interval (CPI) \cite{clopperpearson}. The use of the CPI helps to ensure the probability of joint satisfaction with a given confidence level, $1-\upsilon$. This is expressed by Lemma \ref{lemma:PC}, which is recycled from \cite{MowbrayM2022Sccr}. 

\begin{lemma}\label{lemma:PC} 
\textbf{Joint chance constraint satisfaction via the Clopper-Pearson confidence interval} \cite{clopperpearson}: Consider the realisation of $F_{SA}$ based on $n_I$ independently and identically distributed samples. The lower bound of the true value $F_{LB}$ may be defined with a given confidence $1-\upsilon$, such that:

\begin{equation}
    \begin{aligned}
        &\mathbb{P}(F_U \geq F_{LB}) \geq 1 - \upsilon\\
        &F_{LB} = 1 - \text{betainv}(\upsilon, n_I+1-n_IF_{SA}, n_IF_{SA}) 
    \end{aligned}
\end{equation}
where $\text{betainv}(\cdot)$ is the inverse of the Beta cumulative distribution function with parameters $\{n_I+1-n_IF_{SA}\}$ and $\{n_IF_{SA}\}$.
\end{lemma}

\end{document}